\documentclass[aps,amsmath,amssymb,twocolumn,prb,longbibliography]{revtex4-2}
\usepackage{float}
\usepackage{xcolor}
\usepackage{graphicx}
\usepackage{array}
\newcolumntype{P}[1]{>{\centering\arraybackslash}p{#1}}

\makeatletter
\def\hlinewd#1{%
	\noalign{\ifnum0=`}\fi\hrule \@height #1 \futurelet
	\reserved@a\@xhline}
\makeatother

\usepackage[unicode=true,colorlinks=true,urlcolor=blue,citecolor=blue]{hyperref}

\begin{document}
	
	\title{Correlated Electron Effects in Chromium Trihalide Hetostructures with Graphene: A Tight-Binding Model Perspective}
	
	\author{Igor~Rozhansky}
	\affiliation{National Graphene Institute, University of Manchester, Manchester M13 9PL, United Kingdom} 
	\author{Vladimir Fal'ko}
	\affiliation{National Graphene Institute, University of Manchester, Manchester M13 9PL, United Kingdom} 
	
	\email{igor.rozhanskiy@manchester.ac.uk}

	\begin{abstract} 
In this study, we present an effective tight-binding model 
for an accurate description of the lowest energy quadruplet of conduction band in a ferromagnetic  CrX$_3$ monolayer, tuned to the complementary \textit{ab initio} density functional theory simulations.
This model, based on a minimum number of chromium orbitals,  captures a  distinctively flat dispersion in those bands but requires taking into account hoppings beyond nearest neighbours, revealing ligand-mediated electron pathways connecting remote chromium sites.
Doping of states in the lowest conduction band of CrX$_3$ requires 
charge transfer, which, according to recent studies~\cite{Morpurgo2022,doi:10.1021/acs.nanolett.2c02931,MacDonald2023}, can occur in graphene(G)/CrX$_3$ heterostructures. Here, we use  
the detailed description of the lowest conduction band in CrI$_3$ to show that 
G/CrI$_3$/G and G/CrI$_3$ are type-II heterostructures where light holes in graphene would coexist with heavy electrons in 
the magnetic layer, where the latter can be characterised by Wigner parameter $r_s\sim 15-20$ (as estimated for hBN-encapsulated structures).

		\end{abstract}

	\date{\today}
	\maketitle
	
\section{Introduction}
Chromium trihalides (CrX$_3$, where X = Cl, Br, or I) form a fascinating family of van der Waals materials, celebrated for their versatile magnetic properties~\cite{Kurebayashi2022,Huang2020,Soriano2020,Burch2018,Wang2022,Gibertini2019,McGuire2015,https://doi.org/10.1002/andp.201900452,doi:10.1126/science.aav6926}. 
Over the recent years a broad range of studies of these magnetic insulators has been performed on both bulk materials and atomically thin films produced by mechanical exfoliation~\cite{Morpurgo2022,Morpurgo2023,Yang2021,PhysRevB.100.205409,PhysRevB.106.134412,10.1063/5.0074848,Katsnelson2021,Katsnelson2020,Louie2022,LouieNature2019,Akram2021,Beck2021,Fumega_2023,Xie2022,Dolui2020}.
While the main focus of those  studies was on magnetic properties of CrX$_3$ compounds and their dependence on the number of layers ~\cite{FalkoSong2022,Song2023,SORIANO2019,Beck2022,PhysRevB.105.L081104,Morpurgo2023,SORIANO2019,Song2023,Klein2019,PhysRevB.98.144411}, various CrX$_3$ films were also implemented in heterostructures with other two-dimensional (2D) materials, like graphene, with a view to proximitise ferromagnetic exchange~\cite{MacDonald2023,doi:10.1021/acs.nanolett.2c02931,Zhong2020,Wu2021}. 
 A by-product of such studies was an observation of a substantial charge  transfer between graphene and CrX$_3$ reported by several groups~\cite{Morpurgo2022,doi:10.1021/acs.nanolett.2c02931}, attributed to electrons filling  narrow  conduction bands of  
 CrX$_3$~\cite{Katsnelson2021, Louie2022}, rather than impurity states inside its bandgap. 
 
\begin{figure}
	\centering
	\includegraphics[width=0.5\textwidth]{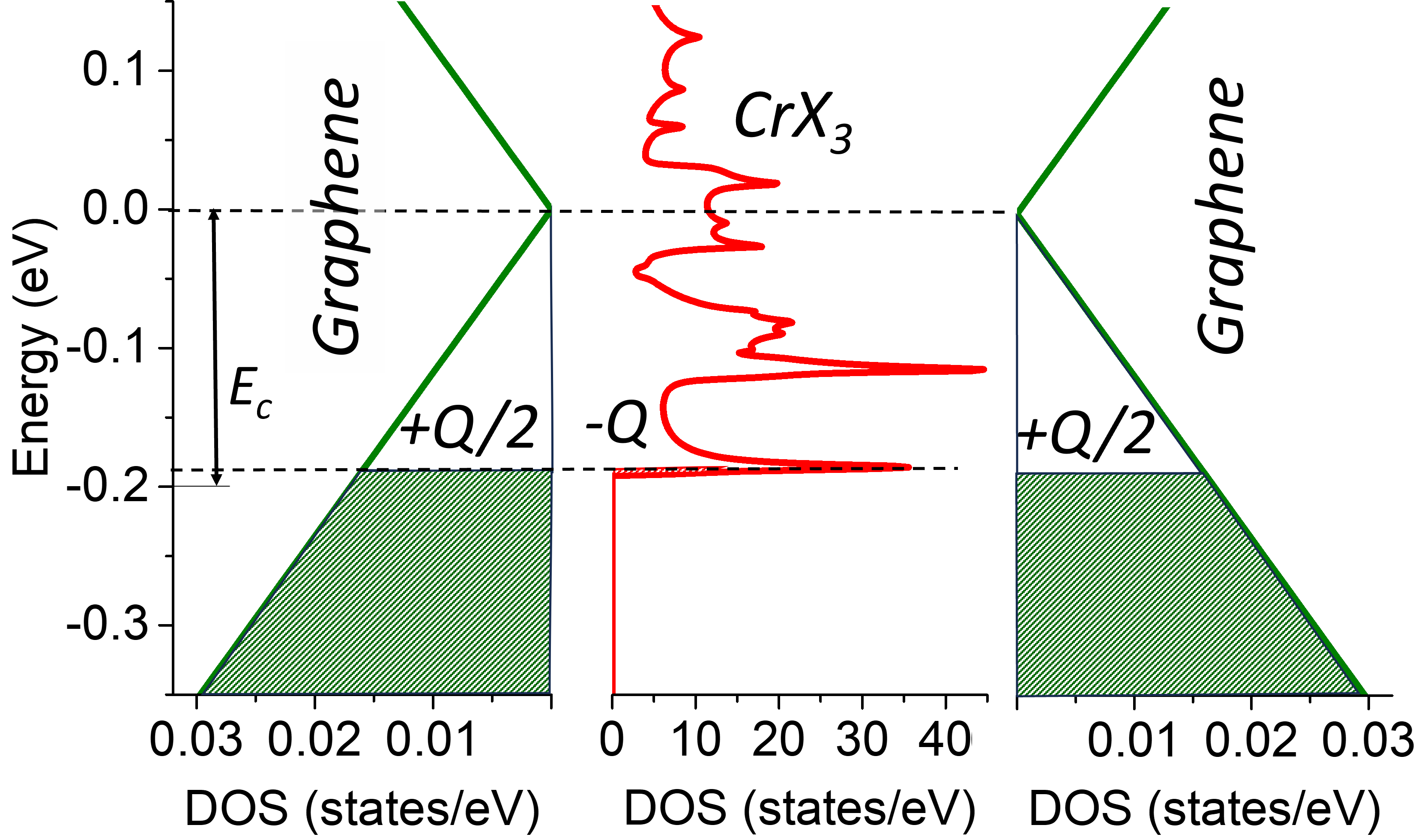}%
	\caption{{\bf Charge transfer in G/CrX$_3$/G trilayer}. The density of states for graphene (left and right) and CrX$_3$ (center) is shown with an offset $E_c$ ($E_c=0.2$ eV for CrI$_3$ as in Ref.~\cite{Zhang2018}), which leads to the transfer of electrons from graphene to the lowest empty band of the d-orbitals-based quadruplet in CrX$_3$ highlighted in Fig.~\ref{figDFT}. Painted  areas indicate occupied states.}
	\label{fig1}
\end{figure}

 The above-mentioned observation opens an interesting avenue towards creating a 2D material that would combine both highly mobile holes in graphene with strongly-correlated heavy electrons in CrX$_3$. Such a system is sketched in Fig.~\ref{fig1}: a trilayer assembled from two graphenes with an embedded CrX$_3$ monolayer, where the transfer of electrons is hosted by  
the lowest spin-polarised conduction band of  CrX$_3$. To describe this band (together with three more bands that belong to a quadruplet traced~\cite{Wu2019,Louie2022,Georgescu2022,FalkoSong2022,Katsnelson2021,https://doi.org/10.1002/adma.202209513,C8CP07067A} to d-orbitals of chromium), we develop an effective tight-binding (TB) model based on a minimal number of Cr orbitals and parametrised by comparison with density-functional theory (DFT) calculations for CrI$_3$. This gives us an access to the accurate description of the conduction band edge across the entire Brillouin zone, hence, obtaining a description of doping features of G/CrI$_3$/G stacks, and a possibility to estimate the Wigner parameter for the heavy electrons.

\begin{figure*}
	\centering
		\includegraphics[width=0.9\textwidth]{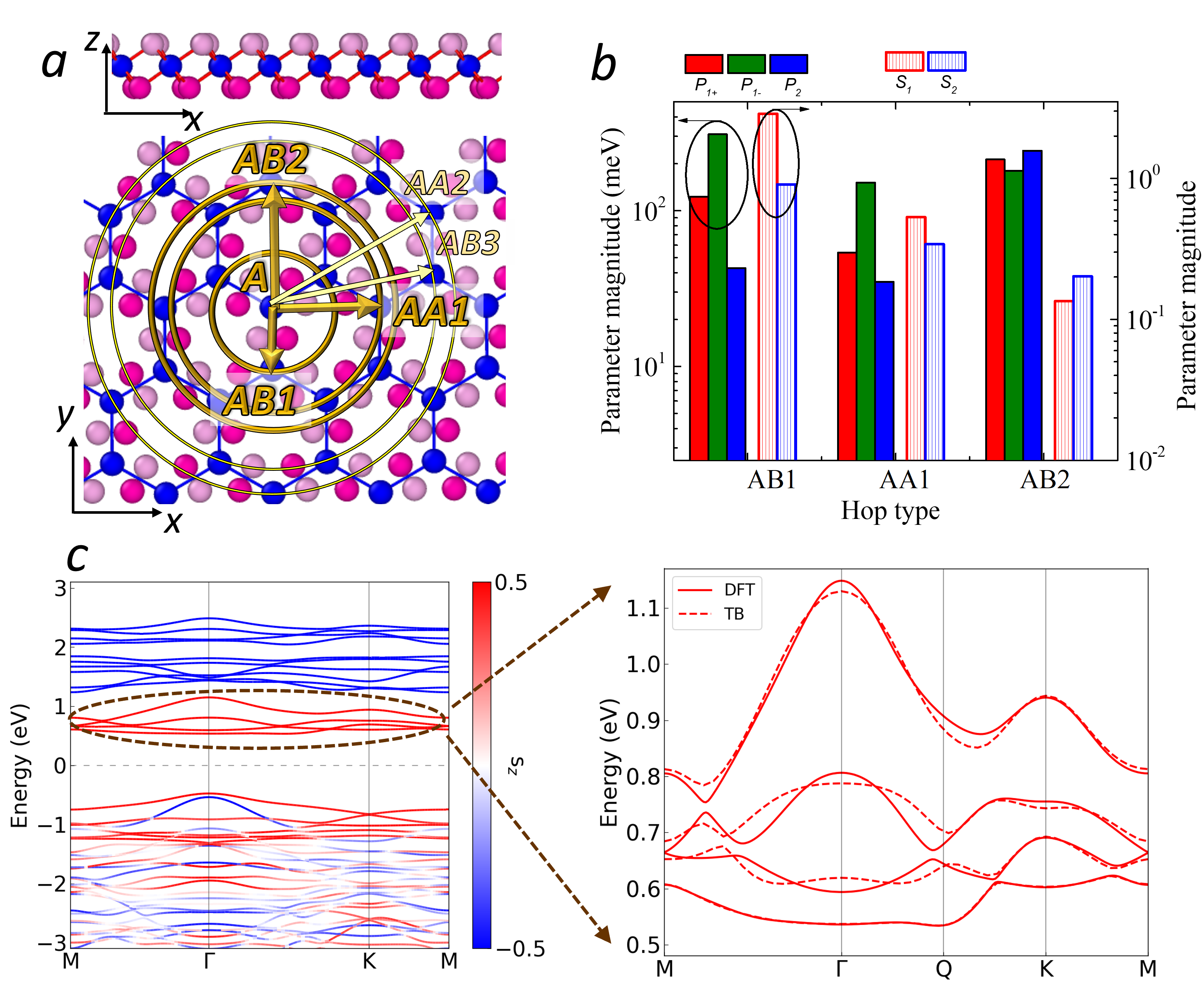}%
				\caption{{\bf Parametrisation of the tight-binding model of CrI$_3$ using its DFT-computed band structure.} (a) Lattice structure of CrX$_3$ monolayers  (X=I,Cl,Br). Cr ions (blue circles) form a honeycomb lattice; lighter and darker pink circles show halogen positions in upper and lower sublayers. Yellow lines show coordination circles labelled with a hop type.
			(b) The magnitudes of parameters used in the parametrisation.		
			(c) Left: DFT-calculated band structure of CrI$_3$ (with SOC taken into account) with colour-coding reflecting spin polarisation of the bands. The bottom of conduction band is dominated by a quadruplet formed of chromium orbitals in Eq.~\ref{eqBasis} implemented in the tight-binding model~(\ref{eqtbham}). 
			Right: the result of tight-binding model~(\ref{eqtbham}) parametrisation (Table~\ref{tabSK}), which accounts for Cr-Cr hoppings up to 3rd neighbour, compared with the DFT data.}
		\label{figDFT}
	\end{figure*} 
\section{Tight-binding model for CrX$_3$ monolayer}
In the crystal structure~\cite{Ubrig_2020,LiuGuoChenGongLiNiuChengLuDengPeng+2022+4409+4417,Li2019,PhysRevB.98.104307} of CrX$_3$,
illustrated in Fig.~\ref{figDFT}a,  metal ions form a honeycomb middle-sublayer lattice, bonded by halogen atoms in the outer sublayers. Following an analogy with graphene, we 
divide  Cr sites into $A$ and $B$ sublattices. Similarly, to graphene, the crystal lattice has inversion symmetry with respect to the centres of the honeycombs and $y\rightarrow -y$ mirror symmetry; however, it lacks mirror symmetry with respect to the horizontal plane and $x\rightarrow -x$ mirror symmetry, due to three out of six nearest  halogens atoms lifted into the top sublayer and the other three pushed down to the bottom sublayers. This makes the symmetry group of the crystal D3d.    

Recent \textit{ab initio} DFT modelling of  
various CrX$_3$ monolayers has 
indicated that wavefunctions of a quadruplet of the lowest spin-polarized conduction bands are dominated by d-orbitals of chromium atoms~\cite{Wu2019,Louie2022,Georgescu2022,FalkoSong2022,Katsnelson2021,https://doi.org/10.1002/adma.202209513,C8CP07067A}. 
Therefore, in the TB model described below, we  
implement a basis of mixed orbitals that would belong to $e_{g\sigma}$ doublet at each Cr site, where, in addition to d-orbitals we include a permitted admixture of p-orbitals as a way to mimic the hybridisation with halogen atoms: 
\begin{align}
	\label{eqBasis}
&{\psi _{A1}} = \alpha Y_2^2 - \beta \left( {\zeta Y_2^{ - 1} + \eta Y_1^{ - 1}} \right);
\nonumber\\
&{\psi _{A2}} = \alpha Y_2^{ - 2} + \beta \left( {\zeta Y_2^1 - \eta Y_1^1} \right); 
\nonumber\\
&{\psi _{B1}} = \alpha Y_2^2 - \beta \left( {\zeta Y_2^{ - 1} - \eta Y_1^{ - 1}} \right);
  \nonumber\\
&{\psi _{B2}} = \alpha Y_2^{ - 2} + \beta \left( {\zeta Y_2^1 + \eta Y_1^1} \right); 
\nonumber\\
&{\alpha ^2} + {\beta ^2} =1;\quad {\zeta ^2} + {\eta ^2} = 1.
\end{align}
Here, $Y_l^m$ are spherical harmonics; $\alpha,\beta$ are on-site mixing parameters for $m=2(-2)$ and $m=-1(1)$ angular harmonics; $\zeta,\eta$ describe the mixing between $l=1$ and $l=2$ harmonics with $|m|=1$; $\psi_{Aj}$ and $\psi_{Bj}$ are associated with two orbitals ($j=1,2$) at A and B sites, respectively. The spatial distribution of the density of such basis states reflects the 3-fold rotational symmetry of the CrX$_3$ lattice.
This basis gives $E \times E$ reducible representation of the D3d point group.

We use this basis as a minimal set to formulate an effective TB model describing the conduction band quadruplet highlighted in the  CrI$_3$ monolayer band-structure displayed in Fig.~\ref{figDFT}b. 
 This band structure was obtained by DFT modelling of the ferromagnetic CrI$_3$ monolayer (magnetized along $z$-axis) using the DFT+$U$+$J$ scheme within the  Quantum-Espresso \textit{ab initio} package~\cite{Giannozzi_2009,Giannozzi_2017} with fully relativistic pseudopotentials (thus, taking the full account of spin-orbit coupling (SOC)) and the Perdew-Burke-Ernzerhof (PBE) approximation for the exchange-correlation
functional~\cite{PBE}. For $U$ and $J$, we take $U$ = 1.5 eV and $J$ = 0.5 eV~\cite{LouieNature2019}. 

For the plane-wave version of the TB model we use Bloch functions and effective Schrodinger equation, 
\begin{align}
	\label{eqShcr}
	&\chi _{Aj,\bf{q}}({\bf{r}}) = \sum\limits_{\bf{R}} {\frac{{{e^{i{\bf{qR}}}}}}{{\sqrt N }}} {\psi _{Aj}}\left( {{\bf{r}} - {\bf{R}}} \right);
	\nonumber\\ 
	&\chi _{Bj,\bf{q}}({\bf{r}}) = \sum\limits_{\bf{R}} {\frac{{{e^{i{\bf{qR}}}}}}{{\sqrt N }}} {\psi _{Bj}}\left( {{\bf{r}} - {\bf{R}}} \right);
	\nonumber\\ 
&\left( {{{\cal H}_{\mathbf q}} - {E_{\mathbf k}}{{\cal{S}}_{\mathbf q}}} \right){\left( {\begin{array}{*{20}{c}}
			{{A_1}}  \\
			{{A_2}}  \\
			{{B_1}}  \\
			{{B_2}}  \\
	\end{array}} \right)_k} = 0,
\end{align}
where $\mathbf{R}$ are vectors of dimensionless Bravais lattice with a unit period, ${\bf q}={\bf k} a_0$ is dimensionless wavevector (normalised by lattice constant $a_0$, ${\bf k}$ is the wavevector),
 and ${\cal H}_{\bf q}$ and ${\cal S}_{\bf q}$ are  
the TB Hamiltonian and overlap matrix.
We formally describe the structure of ${\cal S}$ and ${\cal H}$ using Slater-Koster approach as follows: 
\begin{widetext}
	\begin{align}
		\label{eqtbham}
&{\cal H}_{\bf{q}}  = \sum\limits_{i,j}^{} {\left( {{{\left( { - 1} \right)}^{\left( {j - 1} \right)}}{\varepsilon _s}{\delta _{ij}} + t_{ij}^{\left( 2 \right)}} \right)} \left( {a_i^ + {a_j} + b_i^ + {b_j}} \right) + \sum\limits_{i,j}^{\lambda  = 1,3} {t_{ij}^{\left( \lambda  \right)}a_i^ + {b_j} + H.c.}; 
	\nonumber\\
&{\cal{S}}_{\bf{q}} = \sum\limits_{i,j} {\left( {{\delta _{ij}} + s_{ij}^{\left( 2 \right)}} \right)\left( {a_i^ + {a_i} + b_i^ + {b_i}} \right) + \sum\limits_{i,j}^{\lambda  = 1,3} {s_{ij}^{\left( \lambda  \right)}a_i^ + {b_j} + H.c.} };
\nonumber\\
&s_{11}^{\left( \lambda  \right)} = s_{22}^{\left( \lambda  \right)} = f_1^{\left( \lambda  \right)}S_1^{\left( \lambda  \right)};\quad s_{12}^{\left( {\lambda  = 1,3} \right)} = f_2^{\left( \lambda  \right)}S_2^{\left( \lambda  \right)};\quad s_{21}^{\left( {\lambda  = 1,3} \right)} = f_3^{\left( \lambda  \right)}S_2^{\left( \lambda  \right)};\quad s_{12}^{\left( 2 \right)} = s_{21}^{\left( 2 \right)*} = f_2^{\left( 2 \right)}S_2^{\left( 2 \right)};
\nonumber\\
&t_{11}^{\left( \lambda  \right)} = f_1^{\left( \lambda  \right)}P_{1 + }^{\left( \lambda  \right)};\,\quad t_{22}^{\left( \lambda  \right)} = f_1^{\left( \lambda  \right)}P_{1 - }^{\left( \lambda  \right)};\quad t_{12}^{\left( {\lambda  = 1,3} \right)} = f_2^{\left( \lambda  \right)}P_2^{\left( \lambda  \right)};\quad t_{21}^{\left( {\lambda  = 1,3} \right)} = f_3^{\left( \lambda  \right)}P_2^{\left( \lambda  \right)};\quad t_{12}^{\left( 2 \right)} = t_{21}^{\left( 2 \right)*} = f_2^{\left( 2 \right)}P_2^{\left( 2 \right)};
\nonumber\\
&S_{n = 1,2}^{\left( \lambda  \right)} = {\alpha ^2}\left( {3S_{dd\sigma }^{\left( \lambda  \right)} + 4S_{dd\pi }^{\left( \lambda  \right)} + S_{dd\delta }^{\left( \lambda  \right)}} \right) - 4{\beta ^2}\left[ {{{\left( { - 1} \right)}^{\lambda \cdot\left( {n - 1} \right)}}{\zeta ^2}\left( {S_{dd\delta }^{\left( \lambda  \right)} - {{\left( { - 1} \right)}^n}S_{dd\pi }^{\left( \lambda  \right)}} \right) + {{\left( { - 1} \right)}^{\lambda \cdot n}}{\eta ^2}\left( {S_{pp\sigma }^{\left( \lambda  \right)} - {{\left( { - 1} \right)}^n}S_{pp\pi }^{\left( \lambda  \right)}} \right)} \right];
\nonumber\\
&P_{1 \pm }^{\left( \lambda  \right)} = \sum\limits_{s = 0,1} {{{\left( { \pm 2} \right)}^s}{\alpha ^2}\left( {3V_{dd\sigma }^{\left( {s,\lambda } \right)} + 4V_{dd\pi }^{\left( {s,\lambda } \right)} + V_{dd\delta }^{\left( {s,\lambda } \right)}} \right) + 4{{\left( { \mp 1} \right)}^s}{\beta ^2}\left[ {{\zeta ^2}\left( {V_{dd\delta }^{\left( {s,\lambda } \right)} + V_{dd\pi }^{\left( {s,\lambda } \right)}} \right) + {{\left( { - 1} \right)}^\lambda }{\eta ^2}\left( {V_{pp\sigma }^{\left( {s,\lambda } \right)} + V_{pp\pi }^{\left( {s,\lambda } \right)}} \right)} \right]} ;
\nonumber\\
&P_2^{\left( \lambda  \right)} = {\alpha ^2}\left( {3V_{dd\sigma }^{\left( {0,\lambda } \right)} - 4V_{dd\pi }^{\left( {0,\lambda } \right)} + V_{dd\delta }^{\left( {0,\lambda } \right)}} \right) - 4{\beta ^2}\left[ {{{\left( { - 1} \right)}^\lambda }{\zeta ^2}\left( {V_{dd\delta }^{\left( {0,\lambda } \right)} - V_{dd\pi }^{\left( {0,\lambda } \right)}} \right) + {\eta ^2}\left( {V_{pp\sigma }^{\left( {0,\lambda } \right)} - V_{pp\pi }^{\left( {0,\lambda } \right)}} \right)}; \right]\nonumber\\
&f_n^{\left( 1 \right)} = \frac{{{e^{ - i{q_y}}}}}{4}\left[ {{e^{\frac{{3i{q_y}}}{2}}}\cos \left( {\frac{{\sqrt 3 {q_x}}}{2} - \frac{{2\pi \left( {n - 1} \right)}}{3}} \right) + \frac{1}{2}} \right];\quad f_n^{\left( 3 \right)} = \frac{{{e^{ - i{q_y}}}}}{4}\left[ {\cos \left( {\sqrt 3 {q_x} + \frac{{2\pi \left( {n - 1} \right)}}{3}} \right) + \frac{{{e^{3i{q_y}}}}}{2}} \right];
\nonumber\\
&f_1^{\left( 2 \right)} = \frac{1}{4}\left( {2\cos \frac{{\sqrt 3 {q_x}}}{2}\cos \frac{{3{q_y}}}{2} + \cos \sqrt 3 {q_x}} \right);\quad f_2^{\left( 2 \right)} = \frac{1}{4}\left( {\cos \sqrt 3 {q_x} - \cos \frac{{\sqrt 3 {q_x}}}{2}\cos \frac{{3{q_y}}}{2} - i\sqrt 3 \sin \frac{{\sqrt 3 {q_x}}}{2}\sin \frac{{3{q_y}}}{2}} \right).	\end{align}
\end{widetext}
Here 
$a_{i=1,2}^{(+)}$  and $b_{i=1,2}^{
(+)}$ are 
projection operators onto $A_{1,2}$ and $B_{1,2}$ components of the 4-spinor in 
Eq.~\ref{eqBasis}; $t^{(\lambda)}_{i,j}$ and 
$s^{(\lambda)}_{i,j}$ are the hopping and overlap parameters coupling $i$ and $j$ orbitals: neighbour rank $\lambda$ identifies the coordination circle for the sites involved in a hop ($\lambda=1$ corresponds to nearest-neighbour hop between different sublattice sites denoted as AB1 in Fig.~\ref{figDFT}a, $\lambda=2$ is the shortest intra-sublattice hop denoted as AA1, $\lambda=3$ is the second-neighbour inter-sublattice hop AB2, \textit{etc}). 
The first term in ${\cal H}$ describes on-site 
splitting of the orbitals due to SOC and account for six shortest ($\lambda=2$) intra-sublattice (A-A and B-B) hops. The second term describes three A-B and B-A hops, taking into 
account both closest ($\lambda=1$) and next-neighbour ($\lambda=3$) processes. 
An advantage of the proposed basis (\ref{eqBasis}) is that each of the hopping elements is factorized into k-dependent functions $f^{(\lambda)}_n(\bf{q})$  and fitting parameters $P^{(\lambda)}_n$. The latter can be formally related to Slater-Koster (SK) parameters~\cite{SlaterKoster} 
for d- and p- orbitals $V^{(s,\lambda)}_\alpha$ of two types indicated by index $s=0,1$. Those marked by  $s=0$ correspond to two-centre SK integrals  of spin-independent time-reversal-symmetric part of the one-electron Hamiltonian of the crystal;  parameters with $s=1$ are  related to the time-reversal symmetry breaking by ferromagnetic ordering brought up by SOC. 
 Similarly, the elements of the overlap matrix ${\cal S}$ are expressed through two-center SK integrals but without any SOC contribution. Further details on 
 hopping parameters are given 
 in Supplemental Material~\cite{supp},  
including a discussion of longer hops (
$\lambda=4,5$).

The fitted values of TB parameters 
are listed in Table~\ref{tabSK} and graphically represented in Fig.~\ref{figDFT}b. 
Figure \ref{figDFT}c displays a direct 
comparison between the lowest spin-polarised conduction bands quadruplets computed  
using DFT (solid lines)
and our TB model (dashed lines). A remarkable feature of the identified parameters is the relevance of hops beyond nearest-neighbour, in particular, the large magnitude of overlap parameters  $S_{1,2}^{(1)}$. 
On the one hand a large inter-site overlap and the non-orthogonality of the basis~(\ref{eqBasis}) would make it difficult to use the proposed TB for many-body calculations. On the other hand, it points to that the hops are mediated by the halogens, which could be used to formulate a TB model with a basis expanded by,   e.g., $p$-orbitals of halogens. This extension of the TB model we leave for future studies and, here, simply use the accurate semi-analytical description of the lowest conduction band in the quadruplet to analyse the charge transfer in CrX$_3$/G heterostructures. 
 \begin{table}[t]
 	\centering
 	 		\begin{tabular}{|P{1.2cm}|P{1.2cm}|P{1.2cm}|P{1.2cm}|P{1.2cm}|P{1.2cm}|}
 		\hline
 		\rule[0pt]{0pt}{2.5ex} 
 		& $P_{1+}^{(\lambda)}$ (meV) & $P_{1-}^{(\lambda)}$ (meV)  & $P_{2}^{(\lambda)}$ (meV)  
 		& $S_1^{(\lambda)}$ & $S_2^{(\lambda)}$ \\
 		\hline
 		\rule[0pt]{0pt}{2.5ex} 
 		$\lambda=1$	& 
 		122 & 31 & -43    & -2.88  &  0.91 \\
 		\hline
 		\rule[0pt]{0pt}{2.5ex} 
 		$\lambda=2$	&{-54}  & {-151}  & -35 &    0.53 &   0.34 \\ 
 		\hline
 		\rule[0pt]{0pt}{2.5ex} 
 		$\lambda=3$	& {213} & {181}  & -243 & -0.13 &  -0.20 \\
 		\hline
 		\multicolumn{3}{|c|}{
 			\rule[0pt]{0pt}{2.5ex} 
 			On-site SOC  $\varepsilon_s$} &
 		\multicolumn{3}{|c|}{
 			\rule[0pt]{0pt}{2.5ex} 
 			38 meV} \\
 		\hline 		
 	\end{tabular}
  	\caption{{\bf The  values 
 			used for the 
 			parametrisation of the tight-binding model}. The rows are for neighbour rank $\lambda$.		
 	 		The precision to which each value is given is based on sensitivity of the discrepancy between the TB and DFT spectra with respect to variation of the corresponding parameter. 
 		 	}
 		 		\label{tabSK}
 \end{table}

\section{Graphene/$\bf{CrI_3}$ heterostructure}
\begin{figure}
	\centering
	\includegraphics[width=0.48\textwidth]{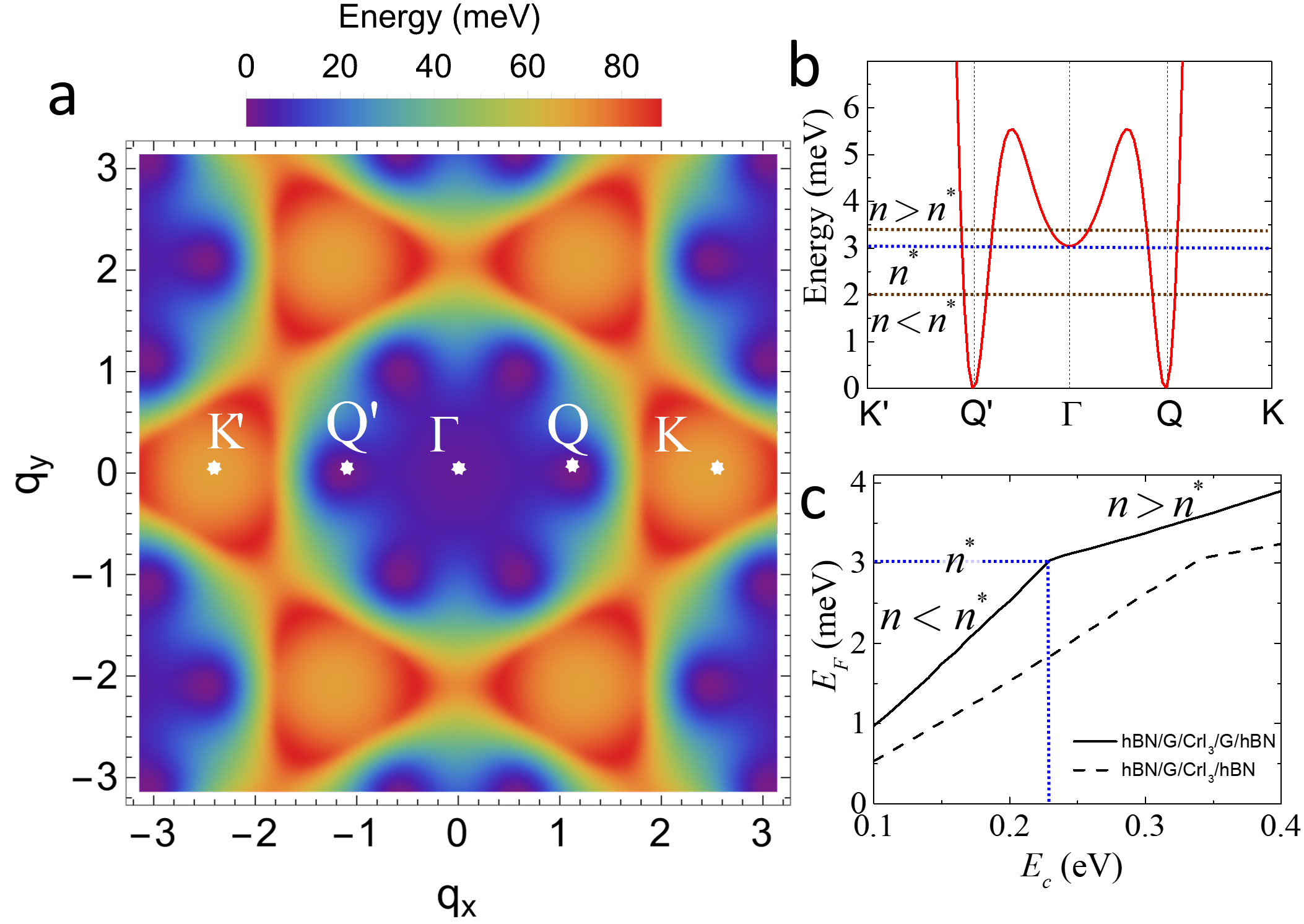}%
	\caption{ {\bf Lower conduction band filling of CrI$_3$ monolayer.}
		(a) The energy profile map of the lower conduction band; the band edge is at six Q,Q' points. 
		(b) Lower energy band profile along $K'-\Gamma - K$ path. Below  critical value of the doping $n^*$ only the Q-point minima are filled, at the doping $n>n^*$ the filling of the minima at $\Gamma$ begins.
		(c) The Fermi level relative to the band edge as a function of G-CrI$_3$ band offset for hBN/G/CrI$_3$/G/hBN (solid line) and hBN/G/CrI$_3$/hBN (dashed line) stacks. 
		The regions corresponding to below critical and above critical value $n^*$ are indicated.} 
	\label{figmass}
\end{figure}

In Fig.~\ref{figDFT}c, the lowest among the conduction band quadruplet of CrI$_3$ is quite flat. The details of the dispersion of this lowest band are elaborated using the semi-analytical description enabled by TB Hamiltonian (\ref{eqtbham}) and are shown in  
Fig~\ref{figmass}a.  
This band features six edge points Q, located approximately halfway along a $\Gamma-K$ path: here we identify three Q and Q' pairs related by time inversion of reciprocal space. 
The dispersion at the band minima is 
parabolic with slightly anisotropic effective masses $m_1=0.54m_0$, $m_2=0.58m_0$, $m_0$ being the free electron mass. An additional minimum at $\Gamma$ point is located $\approx3$ meV above the band edge with an isotropic effective mass exceeding $10m_0$ producing a sufficiently high capacity of electron states for pinning the Fermi level in CrI$_3$ at high doping densities $n>n^*$. From this we conclude that 
in terms of doping the essential bandwidth of CrI$_3$ monolayer is $\approx 3$ meV, so that the charge transfer into it from the environment such as graphene (Fig.~\ref{fig1}) is determined by the interplay between the single-particle band offset, $E_c$, between the above-mentioned CrI$_3$ band and graphene Dirac point, electrostatics (classical capacitance), and Fermi level of holes in graphene (quantum capacitance contribution).
The band edge profile along $K'-\Gamma-K$ direction is shown in Fig~\ref{figmass}c.

It has been reported based on  first-principles calculations that the electron affinity of the CrI$_3$ monolayer exceeds the work function of undoped graphene by $E_c\approx0.2-0.4$ eV, 
suggesting a charge transfer between the layers in G/CrI$_3$ heterostructures~\cite{Zhang2018,MacDonald2023}.
A charge transfer of up to $\sim 10^{13}\, \text{cm}^{-2}$, also electrically tunable 
has been observed experimentally for 
graphene-CrX$_3$ interfaces, 
X=I,Cl,Br~\cite{Morpurgo2022, doi:10.1021/acs.nanolett.2c02931,D2RA02988J,dielectric3,Jiang2019}.
 While one reason for such transfer could be related to defects in the crystal, suggestion have been made that 
the observed graphene p-doping is associated 
with electron transfer into CrX$_3$ conduction band. 
\begin{figure}
	\centering
	\includegraphics[width=0.45\textwidth]{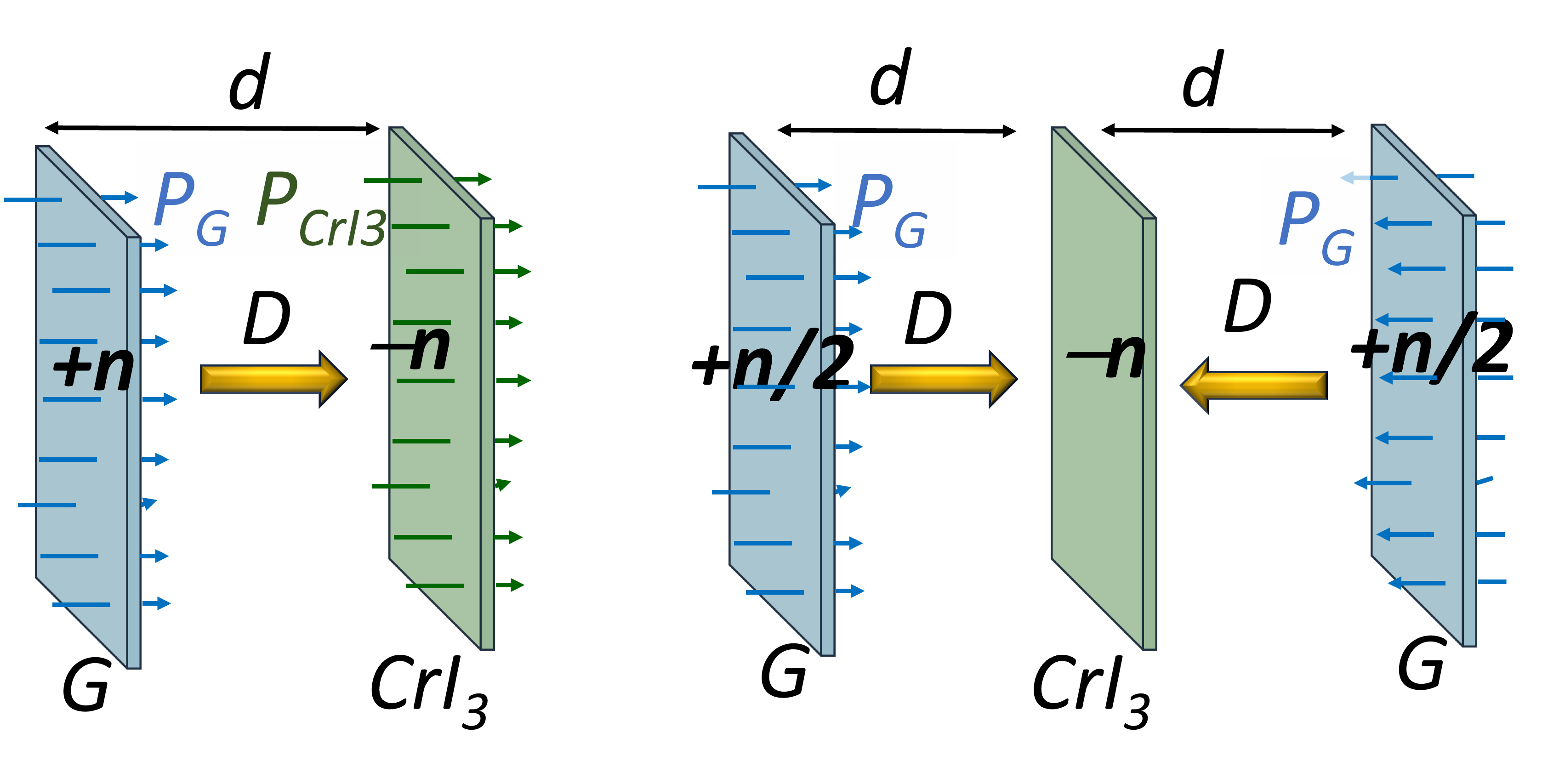}\\
			\includegraphics[width=0.45\textwidth]{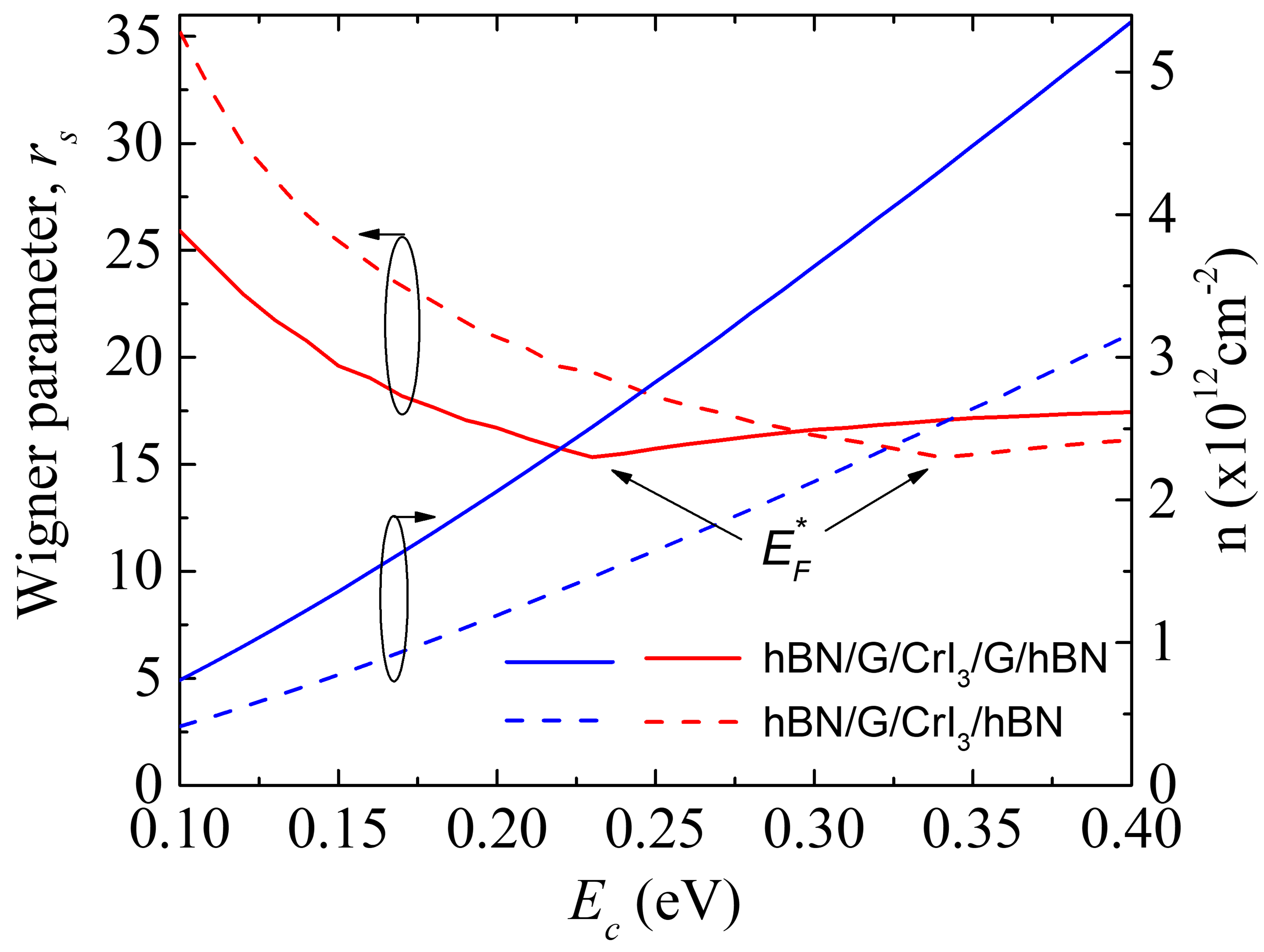}%
	\caption{{\bf Charge transfer in CrI$_3$-graphene stacks.}
			Top: sketches  illustrating charge transfer in a 
			CrI$_3$/G stack (left) and G/CrI$_3$/G (right). The transferred  sheet density $n$ creates a positive charge at graphene and negative charge at CrI$_3$, producing a displacement field $D$ and intra-layer polarisations (see Eqs.~\ref{eqelectro1},\ref{eqelectro2}). 
			Bottom: transferred carrier density, $n$ (r.h.s. axis) and Wigner parameter, $r_s$ (l.h.s. axis) as functions of the band offset, $E_c$. Solid lines correspond to hBN-encapsulated trilayer hBN/G/CrI$_3$/G/hBN, dashed lines correspond to hBN/G/CrI$_3$/hBN. $E_F^*$ corresponds to Fermi level reaching the $\Gamma$ point band edge.   
	}
	\label{figstacks}
\end{figure} 
Using the structure of the lowest conduction band of CrI$_3$ obtained by our modelling, we calculate the charge 
transfer for G/CrI$_3$/G  and G/CrI$_3$ stacks 
shown schematically in Fig.~\ref{figstacks}.

For G/CrI$_3$ stack, the corresponding capacitor model is sketched in Fig.~\ref{figstacks} (left). Here 
a difference $V$ between 
 the on-layer potential energies of graphene and CrI$_3$ layers can be expressed as 
 \begin{align}
 	\label{eqelectro1}
V = \frac{{ed}}{{{\varepsilon _0}}}\left( {en - \frac{{{P_G}}}{2} - \frac{{{P_{{\rm{Cr}}{{\rm{I}}_{\rm{3}}}}}}}{2}} \right);\,\,P = \left( {\varepsilon  - 1} \right)\left( {\frac{{en}}{2} - P} \right),
 \end{align}
  where $n$ is the sheet density of transferred  carriers, $d$ is the distance between the middle of the layers (i.e. carbon and chromium planes), $\varepsilon_{\rm G,CrI_3}$ are the out-of-plane dielectric susceptibilities of graphene and CrI$_3$ monolayers, $n$ is the mobile charge sheet densities in the monolayers, $P_{\rm G,CrI_3}$ denote polarisations of the corresponding layers (see Fig.~\ref{figstacks}a). 
 In the expression for V (Eq.~\ref{eqelectro1}) the first term in brackets  
 is due to displacement field produced by the transferred charge in the layers; while two other terms take into account the 
 layer's dielectric polarisations, as sketched in Fig.~\ref{figstacks}. Here we note that the out-of-plane polarisation of each layer is  caused by the other layer,  which is reflected by a factor of 1/2 in the expression for $P$ in (Eq.~\ref{eqelectro1}) (because the dielectric polarisation of a monolayer is not affected by displacement field generated by transferred charge of itself due to mirror reflection symmetry in each  monolayer plane~\cite{Sliz1}).

For the case of CrI$_3$ sandwiched between two graphene layers, the corresponding capacitor model is sketched in Fig.~\ref{figstacks} (right). In this case, 
the CrI$_3$ layer in the middle is not polarised, due to the symmetry of the structure. 
The expression for the electrostatic energy difference between a graphene and CrI$_3$ monolayer takes the form:
\begin{equation}
\label{eqelectro2}
V = \frac{{ed}}{{{\varepsilon _0}}}\left( {en - \frac{{{P_{\rm G}}}}{2}} \right);\,\,{P_{\rm G}} = \left( {{\varepsilon _{\rm G}} - 1} \right)\left( {\frac{{en}}{2} - {P_{\rm G}}} \right),
\end{equation}
where $n$ is determined as electron density in CrI$_3$ (therefore, both graphene layers are doped with $n/2$ holes).
 
 The carrier density in CrI$_3$, $n$, and the position of the Fermi level for both one- and two-graphene cases can be obtained from the overall charge neutrality of a stack, combined with Eqs.~\ref{eqelectro1},\ref{eqelectro2}:
 \begin{align}
 	\label{eqgr}
 	&n = \frac{{{\varepsilon _0}\alpha V}}{{{e^2}d}} = {N_G}\int\limits_0^{{E_c} - V - {E_F}} {\frac{{2EdE}}{{\pi {v}^2{\hbar ^2}}}}  = \int\limits_0^{{E_F}} \rho \left( E \right)dE, \\	
 	&{\alpha _{{\rm{G/Cr}}{{\rm{I}}_{\rm{3}}}}} = \frac{{4{\varepsilon _{\rm{G}}}{\varepsilon _{{\rm{Cr}}{{\rm{I}}_{\rm{3}}}}}}}{{2{\varepsilon _{\rm{G}}}{\varepsilon _{{\rm{Cr}}{{\rm{I}}_{\rm{3}}}}} + {\varepsilon _{{\rm{Cr}}{{\rm{I}}_{\rm{3}}}}} + {\varepsilon _{\rm{G}}}}};\,\, {\alpha _{{\rm{G/Cr}}{{\rm{I}}_{\rm{3}}}{\rm{/G}}}} = \frac{{8{\varepsilon _{\rm{G}}}}}{{3{\varepsilon _{\rm{G}}} + 1}},
 	\nonumber
 \end{align}
 were  
 $\rho$ is the density of states of the CrI$_3$ monolayer, and $N_G=1,2$ indicates the number of graphene layers in the stack.
 For numerical simulations we use 
 $\varepsilon_{\rm G}=2.6$ ~\cite{Sliz1},    $\varepsilon_{\rm CrI_3}=4$  \cite{LouieNature2019}, and $d=0.5$ nm~\cite{Farooq2019}.   
  The resulting dependence of  $E_F$ and $n$ on the G/Cr$I_3$ band offset, $E_c$,  is 
  shown in Figs.~\ref{figmass}c and \ref{figstacks}, respectively.
  The analysed interval of $E_c$ covers the values suggested in Ref.~\cite{Zhang2018}.
  We note that density $n^*$ corresponds to  $E_c^*\approx0.2$ eV. 
   
    To assess how heavy are the electrons populating the CrI$_3$ band, we compute the Wigner parameter for the analysed interval of doping, Fig.~\ref{figstacks}. Unlike electrostatics determining the charge transfer discussed above, the Coulomb interaction within the layer largely depends on the in-plane polarisation properties of the media and encapsulation of the stacks, in particular 
hBN-encapsulated structures.  
For a bulk hexagonal BN
encapsulation 
$\varepsilon_{\rm{hBN}}=\sqrt{\varepsilon_{\rm{hBN}}^{\parallel c}\varepsilon_{\rm{hBN}}^{\perp c}
}$, where we take $\varepsilon_{\rm{hBN}}^{\parallel c}=3.5$, $\varepsilon_{\rm{hBN}}^{\perp c}=6.9$ for the dielectric constant parallel and perpendicular to c axes, respectively~\cite{Laturia2018,Pierret_2022,PhysRevB.63.115207}, resulting in $\varepsilon_{\rm{hBN}}=4.9$.
Then, we note that,  due to spin+valley degeneracy, number of graphene layers (top and bottom), and steep dispersion of Dirac holes, screening of Coulomb repulsion of heavy electrons in  magnetic layer by graphene is inefficient (in particular, because the average distance between the electrons in CrI$_3$ is much smaller than the Fermi wavelength of  holes in graphene). Therefore, we determine the Wigner parameter as   $$
{r_s} = \frac{{{e^2}\sqrt n }}{{4\pi \varepsilon_{\rm hBN} {\varepsilon _0}{E_F}}}, 
$$
and plot its computed values  in Fig.~\ref{figstacks}. For the range of 
G/CrX$_3$
band offsets 
suggested in the recent literature 
the resulting $r_s$ values fall in the range $15<r_s<20$, which indicate that heavy 
electrons in CrI$_3$ would be strongly correlated.
As a result, here, we get a type-II semiconductor system hosting light highly-mobile holes
in graphene(s) compensating heavy electrons in CrI$_3$, with a Wigner parameter almost in the range of  Wigner crystallisation conditions. 

\section{Discussion}
\begin{figure}
	\centering
	\includegraphics[width=0.5\textwidth]{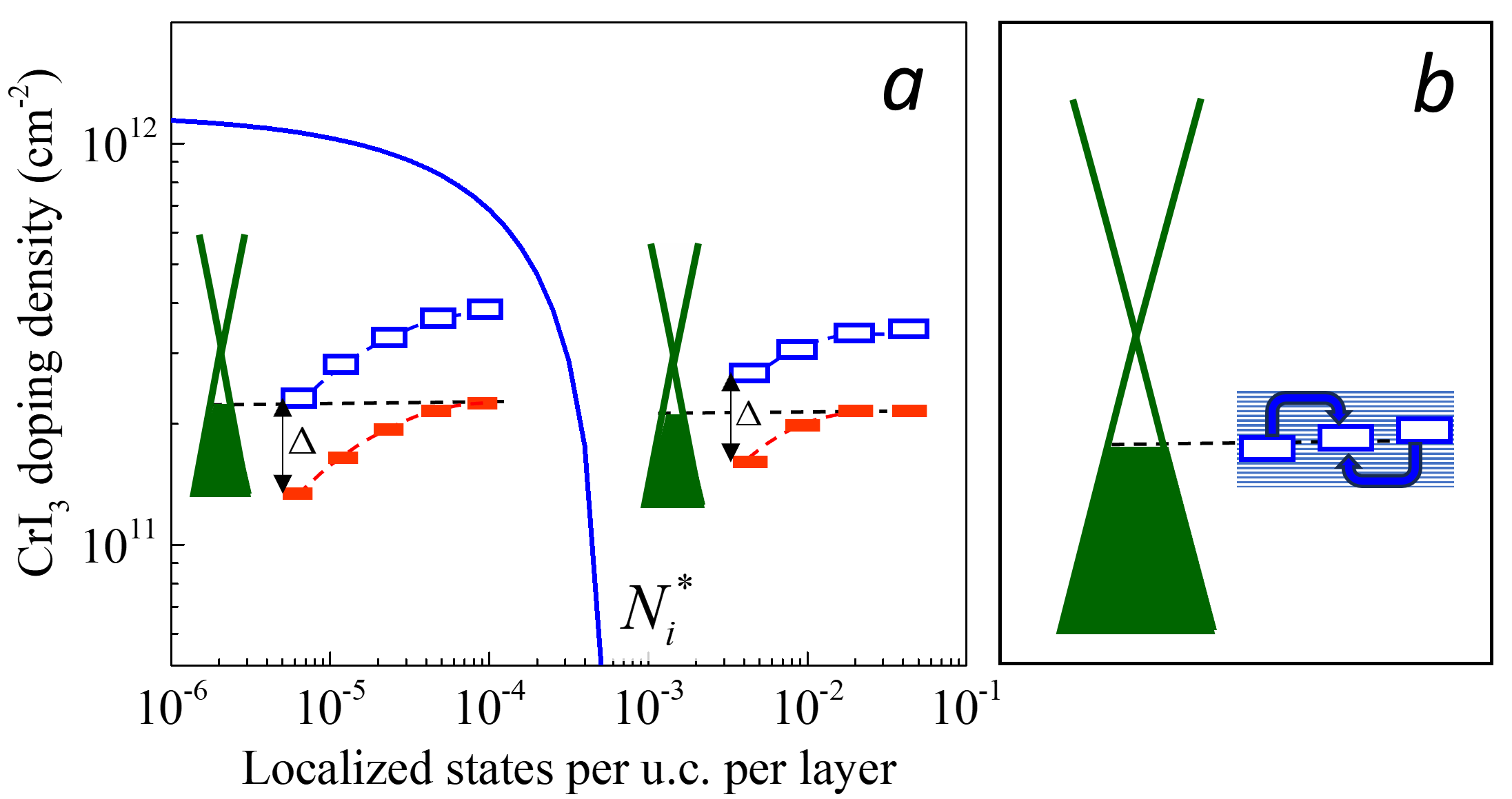}\\
	\caption{{\bf Sketch of a charge transfer between graphene and bulk CrI$_3$}. 
		(a) In-gap defects ($\Delta\sim 0.5$ eV) in multiple CrI$_3$ layers reduce conduction band filling, completely quenching it when the number of localized states per unit cell in each monolayer is $N_i>N_i^*$, as computed for $E_c=0.2$ eV.
		(b) CrI$_3$ conduction band broadening by the interlayer hybridization  could also shrink the bandgap for a bulk material, hence, increasing a charge transfer from graphene.     
		}
	\label{figimp}
\end{figure} 
Based on the data presented in 
   Fig.~\ref{figstacks} we note that the charge transfer to CrI$_3$ monolayer band and the corresponding p-doping of graphene  
   appears to be several times lower than the charger transfer evidenced by  $\sim (5-10)\cdot10^{12}$ cm$^{-2}$ graphene doping (both monolayer and bilayer) experimentally observed in one-sided-graphene on thick CrI$_3$  heterostructures~\cite{Morpurgo2022,doi:10.1021/acs.nanolett.2c02931}. This discrepancy can be attributed to either a contribution of defects producing in-gap states~\cite{doi:10.1021/acs.jpclett.1c00112}, or to a substantial broadening of the conduction band in multi-layer CrI$_3$ by the interlayer hybridisation (thus shrinking the bandgap, similar to what has been found in InSe~\cite{PhysRevB.94.245431, Bandurin2017}). 
   
   We illustrate the influence of in-gap states in~ Fig.\ref{figimp}a by analysing a simultaneous filling of in-gap  and band states for G/CrI$_3$ offset $E_c=0.2$ eV and note that $N_i^*\sim0.5\cdot10^{-3}$ defects per CrI$_3$ unit cell per layer would quench the conduction band occupancy by electron transfer into several CrI$_3$ layers near its surface. From this point of view, the suggested heterostructures based on monolayer CrI$_3$ would reduce the role of in-gap defects. Moreover, graphene-encapsulated CrI$_3$ monolayer (G/CrI$_3$/G) would be an even more promising system for bringing up the above-discussed strongly correlated heavy electrons.

As to the alternative option, related to the interlayer hybridization of CrI$_3$ band states, its careful quantitative description would require an extension of TB model~(\ref{eqtbham}) onto multi-layer structures. For  developing such an extension it is worth noting a dominant role of the next-unit-cell hops between chromium orbital highlighted in Fig.~\ref{figDFT}b. The latter feature of the developed TB model can be attributed to involving of halogen orbitals as intermediate states in the hopping process. At the same time halogen orbitals should be accounted for in the interlayer hybridisation as the halogen ions are located in the outer sub-layers of the 2D crystal. This suggests an alternative formulation of a TB model for CrX$_3$ where the minimum number of Cr orbitals coupled by several neighbour hops will be superseded by a combination of Cr and halogen orbitals with only Cr-halogen and nearest-neighbour halogen-halogen hops (both within and between the  layers).

\section{Acknowledgements}
   We thank A.~Geim, R.~Gorbachev, A.~Morpurgo, A.~Principi, M.~Yankowitz  for stimulating discussions.	
   We acknowledge support
   from EU Graphene Flagship Project, EPSRC Grants No. EP/S019367/1, No. EP/P026850/1, and No. EP/N010345/1,
   and EPSRC CDT Graphene-NOWNANO EP/L01548X/1,
   I.R. gratefully acknowledges the support of British Academy CARA Fellowship.
    
\bibliography{crx}

\begin{thebibliography}{60}%
\makeatletter
\providecommand \@ifxundefined [1]{%
 \@ifx{#1\undefined}
}%
\providecommand \@ifnum [1]{%
 \ifnum #1\expandafter \@firstoftwo
 \else \expandafter \@secondoftwo
 \fi
}%
\providecommand \@ifx [1]{%
 \ifx #1\expandafter \@firstoftwo
 \else \expandafter \@secondoftwo
 \fi
}%
\providecommand \natexlab [1]{#1}%
\providecommand \enquote  [1]{``#1''}%
\providecommand \bibnamefont  [1]{#1}%
\providecommand \bibfnamefont [1]{#1}%
\providecommand \citenamefont [1]{#1}%
\providecommand \href@noop [0]{\@secondoftwo}%
\providecommand \href [0]{\begingroup \@sanitize@url \@href}%
\providecommand \@href[1]{\@@startlink{#1}\@@href}%
\providecommand \@@href[1]{\endgroup#1\@@endlink}%
\providecommand \@sanitize@url [0]{\catcode `\\12\catcode `\$12\catcode
  `\&12\catcode `\#12\catcode `\^12\catcode `\_12\catcode `\%12\relax}%
\providecommand \@@startlink[1]{}%
\providecommand \@@endlink[0]{}%
\providecommand \url  [0]{\begingroup\@sanitize@url \@url }%
\providecommand \@url [1]{\endgroup\@href {#1}{\urlprefix }}%
\providecommand \urlprefix  [0]{URL }%
\providecommand \Eprint [0]{\href }%
\providecommand \doibase [0]{https://doi.org/}%
\providecommand \selectlanguage [0]{\@gobble}%
\providecommand \bibinfo  [0]{\@secondoftwo}%
\providecommand \bibfield  [0]{\@secondoftwo}%
\providecommand \translation [1]{[#1]}%
\providecommand \BibitemOpen [0]{}%
\providecommand \bibitemStop [0]{}%
\providecommand \bibitemNoStop [0]{.\EOS\space}%
\providecommand \EOS [0]{\spacefactor3000\relax}%
\providecommand \BibitemShut  [1]{\csname bibitem#1\endcsname}%
\let\auto@bib@innerbib\@empty
%</preamble>
\bibitem [{\citenamefont {Tenasini}\ \emph {et~al.}(2022)\citenamefont
  {Tenasini}, \citenamefont {Soler-Delgado}, \citenamefont {Wang},
  \citenamefont {Yao}, \citenamefont {Dumcenco}, \citenamefont {Giannini},
  \citenamefont {Watanabe}, \citenamefont {Taniguchi}, \citenamefont
  {Moulsdale}, \citenamefont {Garcia-Ruiz}, \citenamefont {Fal’ko},
  \citenamefont {Gutiérrez-Lezama},\ and\ \citenamefont
  {Morpurgo}}]{Morpurgo2022}%
  \BibitemOpen
  \bibfield  {author} {\bibinfo {author} {\bibfnamefont {G.}~\bibnamefont
  {Tenasini}}, \bibinfo {author} {\bibfnamefont {D.}~\bibnamefont
  {Soler-Delgado}}, \bibinfo {author} {\bibfnamefont {Z.}~\bibnamefont {Wang}},
  \bibinfo {author} {\bibfnamefont {F.}~\bibnamefont {Yao}}, \bibinfo {author}
  {\bibfnamefont {D.}~\bibnamefont {Dumcenco}}, \bibinfo {author}
  {\bibfnamefont {E.}~\bibnamefont {Giannini}}, \bibinfo {author}
  {\bibfnamefont {K.}~\bibnamefont {Watanabe}}, \bibinfo {author}
  {\bibfnamefont {T.}~\bibnamefont {Taniguchi}}, \bibinfo {author}
  {\bibfnamefont {C.}~\bibnamefont {Moulsdale}}, \bibinfo {author}
  {\bibfnamefont {A.}~\bibnamefont {Garcia-Ruiz}}, \bibinfo {author}
  {\bibfnamefont {V.~I.}\ \bibnamefont {Fal’ko}}, \bibinfo {author}
  {\bibfnamefont {I.}~\bibnamefont {Gutiérrez-Lezama}},\ and\ \bibinfo
  {author} {\bibfnamefont {A.~F.}\ \bibnamefont {Morpurgo}},\ }\bibfield
  {title} {\bibinfo {title} {Band gap opening in bilayer
  graphene-{CrCl3/CrBr3/CrI3} van der waals interfaces},\ }\href
  {https://doi.org/10.1021/acs.nanolett.2c02369} {\bibfield  {journal}
  {\bibinfo  {journal} {Nano Letters}\ }\textbf {\bibinfo {volume} {22}},\
  \bibinfo {pages} {6760} (\bibinfo {year} {2022})},\ \bibinfo {note} {pMID:
  35930625},\ \Eprint
  {https://arxiv.org/abs/https://doi.org/10.1021/acs.nanolett.2c02369}
  {https://doi.org/10.1021/acs.nanolett.2c02369} \BibitemShut {NoStop}%
\bibitem [{\citenamefont {Tseng}\ \emph {et~al.}(2022)\citenamefont {Tseng},
  \citenamefont {Song}, \citenamefont {Jiang}, \citenamefont {Lin},
  \citenamefont {Wang}, \citenamefont {Suh}, \citenamefont {Watanabe},
  \citenamefont {Taniguchi}, \citenamefont {McGuire}, \citenamefont {Xiao},
  \citenamefont {Chu}, \citenamefont {Cobden}, \citenamefont {Xu},\ and\
  \citenamefont {Yankowitz}}]{doi:10.1021/acs.nanolett.2c02931}%
  \BibitemOpen
  \bibfield  {author} {\bibinfo {author} {\bibfnamefont {C.-C.}\ \bibnamefont
  {Tseng}}, \bibinfo {author} {\bibfnamefont {T.}~\bibnamefont {Song}},
  \bibinfo {author} {\bibfnamefont {Q.}~\bibnamefont {Jiang}}, \bibinfo
  {author} {\bibfnamefont {Z.}~\bibnamefont {Lin}}, \bibinfo {author}
  {\bibfnamefont {C.}~\bibnamefont {Wang}}, \bibinfo {author} {\bibfnamefont
  {J.}~\bibnamefont {Suh}}, \bibinfo {author} {\bibfnamefont {K.}~\bibnamefont
  {Watanabe}}, \bibinfo {author} {\bibfnamefont {T.}~\bibnamefont {Taniguchi}},
  \bibinfo {author} {\bibfnamefont {M.~A.}\ \bibnamefont {McGuire}}, \bibinfo
  {author} {\bibfnamefont {D.}~\bibnamefont {Xiao}}, \bibinfo {author}
  {\bibfnamefont {J.-H.}\ \bibnamefont {Chu}}, \bibinfo {author} {\bibfnamefont
  {D.~H.}\ \bibnamefont {Cobden}}, \bibinfo {author} {\bibfnamefont
  {X.}~\bibnamefont {Xu}},\ and\ \bibinfo {author} {\bibfnamefont
  {M.}~\bibnamefont {Yankowitz}},\ }\bibfield  {title} {\bibinfo {title}
  {Gate-tunable proximity effects in graphene on layered magnetic insulators},\
  }\href {https://doi.org/10.1021/acs.nanolett.2c02931} {\bibfield  {journal}
  {\bibinfo  {journal} {Nano Letters}\ }\textbf {\bibinfo {volume} {22}},\
  \bibinfo {pages} {8495} (\bibinfo {year} {2022})},\ \bibinfo {note} {pMID:
  36279401},\ \Eprint
  {https://arxiv.org/abs/https://doi.org/10.1021/acs.nanolett.2c02931}
  {https://doi.org/10.1021/acs.nanolett.2c02931} \BibitemShut {NoStop}%
\bibitem [{\citenamefont {Cardoso}\ \emph {et~al.}(2023)\citenamefont
  {Cardoso}, \citenamefont {Costa}, \citenamefont {MacDonald},\ and\
  \citenamefont {Fern\'andez-Rossier}}]{MacDonald2023}%
  \BibitemOpen
  \bibfield  {author} {\bibinfo {author} {\bibfnamefont {C.}~\bibnamefont
  {Cardoso}}, \bibinfo {author} {\bibfnamefont {A.~T.}\ \bibnamefont {Costa}},
  \bibinfo {author} {\bibfnamefont {A.~H.}\ \bibnamefont {MacDonald}},\ and\
  \bibinfo {author} {\bibfnamefont {J.}~\bibnamefont {Fern\'andez-Rossier}},\
  }\bibfield  {title} {\bibinfo {title} {Strong magnetic proximity effect in
  van der waals heterostructures driven by direct hybridization},\ }\href
  {https://doi.org/10.1103/PhysRevB.108.184423} {\bibfield  {journal} {\bibinfo
   {journal} {Phys. Rev. B}\ }\textbf {\bibinfo {volume} {108}},\ \bibinfo
  {pages} {184423} (\bibinfo {year} {2023})}\BibitemShut {NoStop}%
\bibitem [{\citenamefont {Kurebayashi}\ \emph {et~al.}(2022)\citenamefont
  {Kurebayashi}, \citenamefont {Garcia}, \citenamefont {Khan}, \citenamefont
  {Sinova},\ and\ \citenamefont {Roche}}]{Kurebayashi2022}%
  \BibitemOpen
  \bibfield  {author} {\bibinfo {author} {\bibfnamefont {H.}~\bibnamefont
  {Kurebayashi}}, \bibinfo {author} {\bibfnamefont {J.~H.}\ \bibnamefont
  {Garcia}}, \bibinfo {author} {\bibfnamefont {S.}~\bibnamefont {Khan}},
  \bibinfo {author} {\bibfnamefont {J.}~\bibnamefont {Sinova}},\ and\ \bibinfo
  {author} {\bibfnamefont {S.}~\bibnamefont {Roche}},\ }\bibfield  {title}
  {\bibinfo {title} {Magnetism, symmetry and spin transport in van der waals
  layered systems},\ }\href {https://doi.org/10.1038/s42254-021-00403-5}
  {\bibfield  {journal} {\bibinfo  {journal} {Nature Reviews Physics}\ }\textbf
  {\bibinfo {volume} {4}},\ \bibinfo {pages} {150} (\bibinfo {year}
  {2022})}\BibitemShut {NoStop}%
\bibitem [{\citenamefont {Huang}\ \emph {et~al.}(2020)\citenamefont {Huang},
  \citenamefont {McGuire}, \citenamefont {May}, \citenamefont {Xiao},
  \citenamefont {Jarillo-Herrero},\ and\ \citenamefont {Xu}}]{Huang2020}%
  \BibitemOpen
  \bibfield  {author} {\bibinfo {author} {\bibfnamefont {B.}~\bibnamefont
  {Huang}}, \bibinfo {author} {\bibfnamefont {M.~A.}\ \bibnamefont {McGuire}},
  \bibinfo {author} {\bibfnamefont {A.~F.}\ \bibnamefont {May}}, \bibinfo
  {author} {\bibfnamefont {D.}~\bibnamefont {Xiao}}, \bibinfo {author}
  {\bibfnamefont {P.}~\bibnamefont {Jarillo-Herrero}},\ and\ \bibinfo {author}
  {\bibfnamefont {X.}~\bibnamefont {Xu}},\ }\bibfield  {title} {\bibinfo
  {title} {Emergent phenomena and proximity effects in two-dimensional magnets
  and heterostructures},\ }\href {https://doi.org/10.1038/s41563-020-0791-8}
  {\bibfield  {journal} {\bibinfo  {journal} {Nature Materials}\ }\textbf
  {\bibinfo {volume} {19}},\ \bibinfo {pages} {1276} (\bibinfo {year}
  {2020})}\BibitemShut {NoStop}%
\bibitem [{\citenamefont {Soriano}\ \emph {et~al.}(2020)\citenamefont
  {Soriano}, \citenamefont {Katsnelson},\ and\ \citenamefont
  {Fern{\'a}ndez-Rossier}}]{Soriano2020}%
  \BibitemOpen
  \bibfield  {author} {\bibinfo {author} {\bibfnamefont {D.}~\bibnamefont
  {Soriano}}, \bibinfo {author} {\bibfnamefont {M.~I.}\ \bibnamefont
  {Katsnelson}},\ and\ \bibinfo {author} {\bibfnamefont {J.}~\bibnamefont
  {Fern{\'a}ndez-Rossier}},\ }\bibfield  {title} {\bibinfo {title} {Magnetic
  two-dimensional chromium trihalides: A theoretical perspective},\ }\href
  {https://doi.org/10.1021/acs.nanolett.0c02381} {\bibfield  {journal}
  {\bibinfo  {journal} {Nano Letters}\ }\textbf {\bibinfo {volume} {20}},\
  \bibinfo {pages} {6225} (\bibinfo {year} {2020})}\BibitemShut {NoStop}%
\bibitem [{\citenamefont {Burch}\ \emph {et~al.}(2018)\citenamefont {Burch},
  \citenamefont {Mandrus},\ and\ \citenamefont {Park}}]{Burch2018}%
  \BibitemOpen
  \bibfield  {author} {\bibinfo {author} {\bibfnamefont {K.~S.}\ \bibnamefont
  {Burch}}, \bibinfo {author} {\bibfnamefont {D.}~\bibnamefont {Mandrus}},\
  and\ \bibinfo {author} {\bibfnamefont {J.-G.}\ \bibnamefont {Park}},\
  }\bibfield  {title} {\bibinfo {title} {Magnetism in two-dimensional van der
  waals materials},\ }\href {https://doi.org/10.1038/s41586-018-0631-z}
  {\bibfield  {journal} {\bibinfo  {journal} {Nature}\ }\textbf {\bibinfo
  {volume} {563}},\ \bibinfo {pages} {47} (\bibinfo {year} {2018})}\BibitemShut
  {NoStop}%
\bibitem [{\citenamefont {Wang}\ \emph {et~al.}(2022)\citenamefont {Wang},
  \citenamefont {Bedoya-Pinto}, \citenamefont {Blei}, \citenamefont {Dismukes},
  \citenamefont {Hamo}, \citenamefont {Jenkins}, \citenamefont {Koperski},
  \citenamefont {Liu}, \citenamefont {Sun}, \citenamefont {Telford},
  \citenamefont {Kim}, \citenamefont {Augustin}, \citenamefont {Vool},
  \citenamefont {Yin}, \citenamefont {Li}, \citenamefont {Falin}, \citenamefont
  {Dean}, \citenamefont {Casanova}, \citenamefont {Evans}, \citenamefont
  {Chshiev}, \citenamefont {Mishchenko}, \citenamefont {Petrovic},
  \citenamefont {He}, \citenamefont {Zhao}, \citenamefont {Tsen}, \citenamefont
  {Gerardot}, \citenamefont {Brotons-Gisbert}, \citenamefont {Guguchia},
  \citenamefont {Roy}, \citenamefont {Tongay}, \citenamefont {Wang},
  \citenamefont {Hasan}, \citenamefont {Wrachtrup}, \citenamefont {Yacoby},
  \citenamefont {Fert}, \citenamefont {Parkin}, \citenamefont {Novoselov},
  \citenamefont {Dai}, \citenamefont {Balicas},\ and\ \citenamefont
  {Santos}}]{Wang2022}%
  \BibitemOpen
  \bibfield  {author} {\bibinfo {author} {\bibfnamefont {Q.~H.}\ \bibnamefont
  {Wang}}, \bibinfo {author} {\bibfnamefont {A.}~\bibnamefont {Bedoya-Pinto}},
  \bibinfo {author} {\bibfnamefont {M.}~\bibnamefont {Blei}}, \bibinfo {author}
  {\bibfnamefont {A.~H.}\ \bibnamefont {Dismukes}}, \bibinfo {author}
  {\bibfnamefont {A.}~\bibnamefont {Hamo}}, \bibinfo {author} {\bibfnamefont
  {S.}~\bibnamefont {Jenkins}}, \bibinfo {author} {\bibfnamefont
  {M.}~\bibnamefont {Koperski}}, \bibinfo {author} {\bibfnamefont
  {Y.}~\bibnamefont {Liu}}, \bibinfo {author} {\bibfnamefont {Q.-C.}\
  \bibnamefont {Sun}}, \bibinfo {author} {\bibfnamefont {E.~J.}\ \bibnamefont
  {Telford}}, \bibinfo {author} {\bibfnamefont {H.~H.}\ \bibnamefont {Kim}},
  \bibinfo {author} {\bibfnamefont {M.}~\bibnamefont {Augustin}}, \bibinfo
  {author} {\bibfnamefont {U.}~\bibnamefont {Vool}}, \bibinfo {author}
  {\bibfnamefont {J.-X.}\ \bibnamefont {Yin}}, \bibinfo {author} {\bibfnamefont
  {L.~H.}\ \bibnamefont {Li}}, \bibinfo {author} {\bibfnamefont
  {A.}~\bibnamefont {Falin}}, \bibinfo {author} {\bibfnamefont {C.~R.}\
  \bibnamefont {Dean}}, \bibinfo {author} {\bibfnamefont {F.}~\bibnamefont
  {Casanova}}, \bibinfo {author} {\bibfnamefont {R.~F.~L.}\ \bibnamefont
  {Evans}}, \bibinfo {author} {\bibfnamefont {M.}~\bibnamefont {Chshiev}},
  \bibinfo {author} {\bibfnamefont {A.}~\bibnamefont {Mishchenko}}, \bibinfo
  {author} {\bibfnamefont {C.}~\bibnamefont {Petrovic}}, \bibinfo {author}
  {\bibfnamefont {R.}~\bibnamefont {He}}, \bibinfo {author} {\bibfnamefont
  {L.}~\bibnamefont {Zhao}}, \bibinfo {author} {\bibfnamefont {A.~W.}\
  \bibnamefont {Tsen}}, \bibinfo {author} {\bibfnamefont {B.~D.}\ \bibnamefont
  {Gerardot}}, \bibinfo {author} {\bibfnamefont {M.}~\bibnamefont
  {Brotons-Gisbert}}, \bibinfo {author} {\bibfnamefont {Z.}~\bibnamefont
  {Guguchia}}, \bibinfo {author} {\bibfnamefont {X.}~\bibnamefont {Roy}},
  \bibinfo {author} {\bibfnamefont {S.}~\bibnamefont {Tongay}}, \bibinfo
  {author} {\bibfnamefont {Z.}~\bibnamefont {Wang}}, \bibinfo {author}
  {\bibfnamefont {M.~Z.}\ \bibnamefont {Hasan}}, \bibinfo {author}
  {\bibfnamefont {J.}~\bibnamefont {Wrachtrup}}, \bibinfo {author}
  {\bibfnamefont {A.}~\bibnamefont {Yacoby}}, \bibinfo {author} {\bibfnamefont
  {A.}~\bibnamefont {Fert}}, \bibinfo {author} {\bibfnamefont {S.}~\bibnamefont
  {Parkin}}, \bibinfo {author} {\bibfnamefont {K.~S.}\ \bibnamefont
  {Novoselov}}, \bibinfo {author} {\bibfnamefont {P.}~\bibnamefont {Dai}},
  \bibinfo {author} {\bibfnamefont {L.}~\bibnamefont {Balicas}},\ and\ \bibinfo
  {author} {\bibfnamefont {E.~J.~G.}\ \bibnamefont {Santos}},\ }\bibfield
  {title} {\bibinfo {title} {The magnetic genome of two-dimensional van der
  waals materials},\ }\href {https://doi.org/10.1021/acsnano.1c09150}
  {\bibfield  {journal} {\bibinfo  {journal} {ACS Nano}\ }\textbf {\bibinfo
  {volume} {16}},\ \bibinfo {pages} {6960} (\bibinfo {year}
  {2022})}\BibitemShut {NoStop}%
\bibitem [{\citenamefont {Gibertini}\ \emph {et~al.}(2019)\citenamefont
  {Gibertini}, \citenamefont {Koperski}, \citenamefont {Morpurgo},\ and\
  \citenamefont {Novoselov}}]{Gibertini2019}%
  \BibitemOpen
  \bibfield  {author} {\bibinfo {author} {\bibfnamefont {M.}~\bibnamefont
  {Gibertini}}, \bibinfo {author} {\bibfnamefont {M.}~\bibnamefont {Koperski}},
  \bibinfo {author} {\bibfnamefont {A.~F.}\ \bibnamefont {Morpurgo}},\ and\
  \bibinfo {author} {\bibfnamefont {K.~S.}\ \bibnamefont {Novoselov}},\
  }\bibfield  {title} {\bibinfo {title} {Magnetic 2d materials and
  heterostructures},\ }\href {https://doi.org/10.1038/s41565-019-0438-6}
  {\bibfield  {journal} {\bibinfo  {journal} {Nature Nanotechnology}\ }\textbf
  {\bibinfo {volume} {14}},\ \bibinfo {pages} {408} (\bibinfo {year}
  {2019})}\BibitemShut {NoStop}%
\bibitem [{\citenamefont {McGuire}\ \emph {et~al.}(2015)\citenamefont
  {McGuire}, \citenamefont {Dixit}, \citenamefont {Cooper},\ and\ \citenamefont
  {Sales}}]{McGuire2015}%
  \BibitemOpen
  \bibfield  {author} {\bibinfo {author} {\bibfnamefont {M.~A.}\ \bibnamefont
  {McGuire}}, \bibinfo {author} {\bibfnamefont {H.}~\bibnamefont {Dixit}},
  \bibinfo {author} {\bibfnamefont {V.~R.}\ \bibnamefont {Cooper}},\ and\
  \bibinfo {author} {\bibfnamefont {B.~C.}\ \bibnamefont {Sales}},\ }\bibfield
  {title} {\bibinfo {title} {Coupling of crystal structure and magnetism in the
  layered, ferromagnetic insulator cri3},\ }\href
  {https://doi.org/10.1021/cm504242t} {\bibfield  {journal} {\bibinfo
  {journal} {Chemistry of Materials}\ }\textbf {\bibinfo {volume} {27}},\
  \bibinfo {pages} {612} (\bibinfo {year} {2015})}\BibitemShut {NoStop}%
\bibitem [{\citenamefont {Wang}\ \emph {et~al.}(2020)\citenamefont {Wang},
  \citenamefont {Huang}, \citenamefont {Cheung}, \citenamefont {Chen},
  \citenamefont {Tan}, \citenamefont {Huang}, \citenamefont {Zhao},
  \citenamefont {Zhao}, \citenamefont {Wu}, \citenamefont {Feng}, \citenamefont
  {Wu},\ and\ \citenamefont {Chang}}]{https://doi.org/10.1002/andp.201900452}%
  \BibitemOpen
  \bibfield  {author} {\bibinfo {author} {\bibfnamefont {M.-C.}\ \bibnamefont
  {Wang}}, \bibinfo {author} {\bibfnamefont {C.-C.}\ \bibnamefont {Huang}},
  \bibinfo {author} {\bibfnamefont {C.-H.}\ \bibnamefont {Cheung}}, \bibinfo
  {author} {\bibfnamefont {C.-Y.}\ \bibnamefont {Chen}}, \bibinfo {author}
  {\bibfnamefont {S.~G.}\ \bibnamefont {Tan}}, \bibinfo {author} {\bibfnamefont
  {T.-W.}\ \bibnamefont {Huang}}, \bibinfo {author} {\bibfnamefont
  {Y.}~\bibnamefont {Zhao}}, \bibinfo {author} {\bibfnamefont {Y.}~\bibnamefont
  {Zhao}}, \bibinfo {author} {\bibfnamefont {G.}~\bibnamefont {Wu}}, \bibinfo
  {author} {\bibfnamefont {Y.-P.}\ \bibnamefont {Feng}}, \bibinfo {author}
  {\bibfnamefont {H.-C.}\ \bibnamefont {Wu}},\ and\ \bibinfo {author}
  {\bibfnamefont {C.-R.}\ \bibnamefont {Chang}},\ }\bibfield  {title} {\bibinfo
  {title} {Prospects and opportunities of 2d van der waals magnetic systems},\
  }\href {https://doi.org/https://doi.org/10.1002/andp.201900452} {\bibfield
  {journal} {\bibinfo  {journal} {Annalen der Physik}\ }\textbf {\bibinfo
  {volume} {532}},\ \bibinfo {pages} {1900452} (\bibinfo {year} {2020})},\
  \Eprint
  {https://arxiv.org/abs/https://onlinelibrary.wiley.com/doi/pdf/10.1002/andp.201900452}
  {https://onlinelibrary.wiley.com/doi/pdf/10.1002/andp.201900452} \BibitemShut
  {NoStop}%
\bibitem [{\citenamefont {Thiel}\ \emph {et~al.}(2019)\citenamefont {Thiel},
  \citenamefont {Wang}, \citenamefont {Tschudin}, \citenamefont {Rohner},
  \citenamefont {Gutiérrez-Lezama}, \citenamefont {Ubrig}, \citenamefont
  {Gibertini}, \citenamefont {Giannini}, \citenamefont {Morpurgo},\ and\
  \citenamefont {Maletinsky}}]{doi:10.1126/science.aav6926}%
  \BibitemOpen
  \bibfield  {author} {\bibinfo {author} {\bibfnamefont {L.}~\bibnamefont
  {Thiel}}, \bibinfo {author} {\bibfnamefont {Z.}~\bibnamefont {Wang}},
  \bibinfo {author} {\bibfnamefont {M.~A.}\ \bibnamefont {Tschudin}}, \bibinfo
  {author} {\bibfnamefont {D.}~\bibnamefont {Rohner}}, \bibinfo {author}
  {\bibfnamefont {I.}~\bibnamefont {Gutiérrez-Lezama}}, \bibinfo {author}
  {\bibfnamefont {N.}~\bibnamefont {Ubrig}}, \bibinfo {author} {\bibfnamefont
  {M.}~\bibnamefont {Gibertini}}, \bibinfo {author} {\bibfnamefont
  {E.}~\bibnamefont {Giannini}}, \bibinfo {author} {\bibfnamefont {A.~F.}\
  \bibnamefont {Morpurgo}},\ and\ \bibinfo {author} {\bibfnamefont
  {P.}~\bibnamefont {Maletinsky}},\ }\bibfield  {title} {\bibinfo {title}
  {Probing magnetism in 2d materials at the nanoscale with single-spin
  microscopy},\ }\href {https://doi.org/10.1126/science.aav6926} {\bibfield
  {journal} {\bibinfo  {journal} {Science}\ }\textbf {\bibinfo {volume}
  {364}},\ \bibinfo {pages} {973} (\bibinfo {year} {2019})},\ \Eprint
  {https://arxiv.org/abs/https://www.science.org/doi/pdf/10.1126/science.aav6926}
  {https://www.science.org/doi/pdf/10.1126/science.aav6926} \BibitemShut
  {NoStop}%
\bibitem [{\citenamefont {Yao}\ \emph {et~al.}(2023)\citenamefont {Yao},
  \citenamefont {Multian}, \citenamefont {Wang}, \citenamefont {Ubrig},
  \citenamefont {Teyssier}, \citenamefont {Wu}, \citenamefont {Giannini},
  \citenamefont {Gibertini}, \citenamefont {Guti{\'e}rrez-Lezama},\ and\
  \citenamefont {Morpurgo}}]{Morpurgo2023}%
  \BibitemOpen
  \bibfield  {author} {\bibinfo {author} {\bibfnamefont {F.}~\bibnamefont
  {Yao}}, \bibinfo {author} {\bibfnamefont {V.}~\bibnamefont {Multian}},
  \bibinfo {author} {\bibfnamefont {Z.}~\bibnamefont {Wang}}, \bibinfo {author}
  {\bibfnamefont {N.}~\bibnamefont {Ubrig}}, \bibinfo {author} {\bibfnamefont
  {J.}~\bibnamefont {Teyssier}}, \bibinfo {author} {\bibfnamefont
  {F.}~\bibnamefont {Wu}}, \bibinfo {author} {\bibfnamefont {E.}~\bibnamefont
  {Giannini}}, \bibinfo {author} {\bibfnamefont {M.}~\bibnamefont {Gibertini}},
  \bibinfo {author} {\bibfnamefont {I.}~\bibnamefont {Guti{\'e}rrez-Lezama}},\
  and\ \bibinfo {author} {\bibfnamefont {A.~F.}\ \bibnamefont {Morpurgo}},\
  }\bibfield  {title} {\bibinfo {title} {Multiple antiferromagnetic phases and
  magnetic anisotropy in exfoliated crbr3 multilayers},\ }\href
  {https://doi.org/10.1038/s41467-023-40723-x} {\bibfield  {journal} {\bibinfo
  {journal} {Nature Communications}\ }\textbf {\bibinfo {volume} {14}},\
  \bibinfo {pages} {4969} (\bibinfo {year} {2023})}\BibitemShut {NoStop}%
\bibitem [{\citenamefont {Yang}\ \emph {et~al.}(2021)\citenamefont {Yang},
  \citenamefont {Hu}, \citenamefont {Shen}, \citenamefont {Krasheninnikov},
  \citenamefont {Chen},\ and\ \citenamefont {Sun}}]{Yang2021}%
  \BibitemOpen
  \bibfield  {author} {\bibinfo {author} {\bibfnamefont {Q.}~\bibnamefont
  {Yang}}, \bibinfo {author} {\bibfnamefont {X.}~\bibnamefont {Hu}}, \bibinfo
  {author} {\bibfnamefont {X.}~\bibnamefont {Shen}}, \bibinfo {author}
  {\bibfnamefont {A.~V.}\ \bibnamefont {Krasheninnikov}}, \bibinfo {author}
  {\bibfnamefont {Z.}~\bibnamefont {Chen}},\ and\ \bibinfo {author}
  {\bibfnamefont {L.}~\bibnamefont {Sun}},\ }\bibfield  {title} {\bibinfo
  {title} {Enhancing ferromagnetism and tuning electronic properties of cri3
  monolayers by adsorption of transition-metal atoms},\ }\href
  {https://doi.org/10.1021/acsami.1c01701} {\bibfield  {journal} {\bibinfo
  {journal} {ACS Applied Materials {\&} Interfaces}\ }\textbf {\bibinfo
  {volume} {13}},\ \bibinfo {pages} {21593} (\bibinfo {year}
  {2021})}\BibitemShut {NoStop}%
\bibitem [{\citenamefont {Lu}\ \emph {et~al.}(2019)\citenamefont {Lu},
  \citenamefont {Fei},\ and\ \citenamefont {Yang}}]{PhysRevB.100.205409}%
  \BibitemOpen
  \bibfield  {author} {\bibinfo {author} {\bibfnamefont {X.}~\bibnamefont
  {Lu}}, \bibinfo {author} {\bibfnamefont {R.}~\bibnamefont {Fei}},\ and\
  \bibinfo {author} {\bibfnamefont {L.}~\bibnamefont {Yang}},\ }\bibfield
  {title} {\bibinfo {title} {Curie temperature of emerging two-dimensional
  magnetic structures},\ }\href {https://doi.org/10.1103/PhysRevB.100.205409}
  {\bibfield  {journal} {\bibinfo  {journal} {Phys. Rev. B}\ }\textbf {\bibinfo
  {volume} {100}},\ \bibinfo {pages} {205409} (\bibinfo {year}
  {2019})}\BibitemShut {NoStop}%
\bibitem [{\citenamefont {Zhu}\ \emph {et~al.}(2022)\citenamefont {Zhu},
  \citenamefont {Gao}, \citenamefont {Hou}, \citenamefont {Gui},\ and\
  \citenamefont {Huang}}]{PhysRevB.106.134412}%
  \BibitemOpen
  \bibfield  {author} {\bibinfo {author} {\bibfnamefont {H.}~\bibnamefont
  {Zhu}}, \bibinfo {author} {\bibfnamefont {Y.}~\bibnamefont {Gao}}, \bibinfo
  {author} {\bibfnamefont {Y.}~\bibnamefont {Hou}}, \bibinfo {author}
  {\bibfnamefont {Z.}~\bibnamefont {Gui}},\ and\ \bibinfo {author}
  {\bibfnamefont {L.}~\bibnamefont {Huang}},\ }\bibfield  {title} {\bibinfo
  {title} {Tunable magnetic anisotropy in two-dimensional
  $\mathrm{Cr}{X}_{3}/\mathrm{AlN} (x=\mathrm{I},\mathrm{Br},\mathrm{Cl})$
  heterostructures},\ }\href {https://doi.org/10.1103/PhysRevB.106.134412}
  {\bibfield  {journal} {\bibinfo  {journal} {Phys. Rev. B}\ }\textbf {\bibinfo
  {volume} {106}},\ \bibinfo {pages} {134412} (\bibinfo {year}
  {2022})}\BibitemShut {NoStop}%
\bibitem [{\citenamefont {Staros}\ \emph {et~al.}(2022)\citenamefont {Staros},
  \citenamefont {Hu}, \citenamefont {Tiihonen}, \citenamefont {Nanguneri},
  \citenamefont {Krogel}, \citenamefont {Bennett}, \citenamefont {Heinonen},
  \citenamefont {Ganesh},\ and\ \citenamefont
  {Rubenstein}}]{10.1063/5.0074848}%
  \BibitemOpen
  \bibfield  {author} {\bibinfo {author} {\bibfnamefont {D.}~\bibnamefont
  {Staros}}, \bibinfo {author} {\bibfnamefont {G.}~\bibnamefont {Hu}}, \bibinfo
  {author} {\bibfnamefont {J.}~\bibnamefont {Tiihonen}}, \bibinfo {author}
  {\bibfnamefont {R.}~\bibnamefont {Nanguneri}}, \bibinfo {author}
  {\bibfnamefont {J.}~\bibnamefont {Krogel}}, \bibinfo {author} {\bibfnamefont
  {M.~C.}\ \bibnamefont {Bennett}}, \bibinfo {author} {\bibfnamefont
  {O.}~\bibnamefont {Heinonen}}, \bibinfo {author} {\bibfnamefont
  {P.}~\bibnamefont {Ganesh}},\ and\ \bibinfo {author} {\bibfnamefont
  {B.}~\bibnamefont {Rubenstein}},\ }\bibfield  {title} {\bibinfo {title} {{A
  combined first principles study of the structural, magnetic, and phonon
  properties of monolayer CrI3}},\ }\href {https://doi.org/10.1063/5.0074848}
  {\bibfield  {journal} {\bibinfo  {journal} {The Journal of Chemical Physics}\
  }\textbf {\bibinfo {volume} {156}},\ \bibinfo {pages} {014707} (\bibinfo
  {year} {2022})},\ \Eprint
  {https://arxiv.org/abs/https://pubs.aip.org/aip/jcp/article-pdf/doi/10.1063/5.0074848/16535185/014707\_1\_online.pdf}
  {https://pubs.aip.org/aip/jcp/article-pdf/doi/10.1063/5.0074848/16535185/014707\_1\_online.pdf}
  \BibitemShut {NoStop}%
\bibitem [{\citenamefont {Acharya}\ \emph {et~al.}(2021)\citenamefont
  {Acharya}, \citenamefont {Pashov}, \citenamefont {Cunningham}, \citenamefont
  {Rudenko}, \citenamefont {R\"osner}, \citenamefont {Gr\"uning}, \citenamefont
  {van Schilfgaarde},\ and\ \citenamefont {Katsnelson}}]{Katsnelson2021}%
  \BibitemOpen
  \bibfield  {author} {\bibinfo {author} {\bibfnamefont {S.}~\bibnamefont
  {Acharya}}, \bibinfo {author} {\bibfnamefont {D.}~\bibnamefont {Pashov}},
  \bibinfo {author} {\bibfnamefont {B.}~\bibnamefont {Cunningham}}, \bibinfo
  {author} {\bibfnamefont {A.~N.}\ \bibnamefont {Rudenko}}, \bibinfo {author}
  {\bibfnamefont {M.}~\bibnamefont {R\"osner}}, \bibinfo {author}
  {\bibfnamefont {M.}~\bibnamefont {Gr\"uning}}, \bibinfo {author}
  {\bibfnamefont {M.}~\bibnamefont {van Schilfgaarde}},\ and\ \bibinfo {author}
  {\bibfnamefont {M.~I.}\ \bibnamefont {Katsnelson}},\ }\bibfield  {title}
  {\bibinfo {title} {Electronic structure of chromium trihalides beyond density
  functional theory},\ }\href {https://doi.org/10.1103/PhysRevB.104.155109}
  {\bibfield  {journal} {\bibinfo  {journal} {Phys. Rev. B}\ }\textbf {\bibinfo
  {volume} {104}},\ \bibinfo {pages} {155109} (\bibinfo {year}
  {2021})}\BibitemShut {NoStop}%
\bibitem [{\citenamefont {Kvashnin}\ \emph {et~al.}(2020)\citenamefont
  {Kvashnin}, \citenamefont {Bergman}, \citenamefont {Lichtenstein},\ and\
  \citenamefont {Katsnelson}}]{Katsnelson2020}%
  \BibitemOpen
  \bibfield  {author} {\bibinfo {author} {\bibfnamefont {Y.~O.}\ \bibnamefont
  {Kvashnin}}, \bibinfo {author} {\bibfnamefont {A.}~\bibnamefont {Bergman}},
  \bibinfo {author} {\bibfnamefont {A.~I.}\ \bibnamefont {Lichtenstein}},\ and\
  \bibinfo {author} {\bibfnamefont {M.~I.}\ \bibnamefont {Katsnelson}},\
  }\bibfield  {title} {\bibinfo {title} {Relativistic exchange interactions in
  $\mathrm{Cr}{X}_{3}$ ($x=\mathrm{Cl}$, br, i) monolayers},\ }\href
  {https://doi.org/10.1103/PhysRevB.102.115162} {\bibfield  {journal} {\bibinfo
   {journal} {Phys. Rev. B}\ }\textbf {\bibinfo {volume} {102}},\ \bibinfo
  {pages} {115162} (\bibinfo {year} {2020})}\BibitemShut {NoStop}%
\bibitem [{\citenamefont {Wu}\ \emph {et~al.}(2022)\citenamefont {Wu},
  \citenamefont {Li},\ and\ \citenamefont {Louie}}]{Louie2022}%
  \BibitemOpen
  \bibfield  {author} {\bibinfo {author} {\bibfnamefont {M.}~\bibnamefont
  {Wu}}, \bibinfo {author} {\bibfnamefont {Z.}~\bibnamefont {Li}},\ and\
  \bibinfo {author} {\bibfnamefont {S.~G.}\ \bibnamefont {Louie}},\ }\bibfield
  {title} {\bibinfo {title} {Optical and magneto-optical properties of
  ferromagnetic monolayer ${\mathrm{crbr}}_{3}$: A first-principles $gw$ and
  $gw$ plus bethe-salpeter equation study},\ }\href
  {https://doi.org/10.1103/PhysRevMaterials.6.014008} {\bibfield  {journal}
  {\bibinfo  {journal} {Phys. Rev. Mater.}\ }\textbf {\bibinfo {volume} {6}},\
  \bibinfo {pages} {014008} (\bibinfo {year} {2022})}\BibitemShut {NoStop}%
\bibitem [{\citenamefont {Wu}\ \emph {et~al.}(2019{\natexlab{a}})\citenamefont
  {Wu}, \citenamefont {Li}, \citenamefont {Cao},\ and\ \citenamefont
  {Louie}}]{LouieNature2019}%
  \BibitemOpen
  \bibfield  {author} {\bibinfo {author} {\bibfnamefont {M.}~\bibnamefont
  {Wu}}, \bibinfo {author} {\bibfnamefont {Z.}~\bibnamefont {Li}}, \bibinfo
  {author} {\bibfnamefont {T.}~\bibnamefont {Cao}},\ and\ \bibinfo {author}
  {\bibfnamefont {S.~G.}\ \bibnamefont {Louie}},\ }\bibfield  {title} {\bibinfo
  {title} {Physical origin of giant excitonic and magneto-optical responses in
  two-dimensional ferromagnetic insulators},\ }\href
  {https://doi.org/10.1038/s41467-019-10325-7} {\bibfield  {journal} {\bibinfo
  {journal} {Nature Communications}\ }\textbf {\bibinfo {volume} {10}},\
  \bibinfo {pages} {2371} (\bibinfo {year} {2019}{\natexlab{a}})}\BibitemShut
  {NoStop}%
\bibitem [{\citenamefont {Akram}\ \emph {et~al.}(2021)\citenamefont {Akram},
  \citenamefont {LaBollita}, \citenamefont {Dey}, \citenamefont {Kapeghian},
  \citenamefont {Erten},\ and\ \citenamefont {Botana}}]{Akram2021}%
  \BibitemOpen
  \bibfield  {author} {\bibinfo {author} {\bibfnamefont {M.}~\bibnamefont
  {Akram}}, \bibinfo {author} {\bibfnamefont {H.}~\bibnamefont {LaBollita}},
  \bibinfo {author} {\bibfnamefont {D.}~\bibnamefont {Dey}}, \bibinfo {author}
  {\bibfnamefont {J.}~\bibnamefont {Kapeghian}}, \bibinfo {author}
  {\bibfnamefont {O.}~\bibnamefont {Erten}},\ and\ \bibinfo {author}
  {\bibfnamefont {A.~S.}\ \bibnamefont {Botana}},\ }\bibfield  {title}
  {\bibinfo {title} {Moir{\'e} skyrmions and chiral magnetic phases in twisted
  {CrX3} ({X = I, Br, and Cl}) bilayers},\ }\href
  {https://doi.org/10.1021/acs.nanolett.1c02096} {\bibfield  {journal}
  {\bibinfo  {journal} {Nano Letters}\ }\textbf {\bibinfo {volume} {21}},\
  \bibinfo {pages} {6633} (\bibinfo {year} {2021})}\BibitemShut {NoStop}%
\bibitem [{\citenamefont {Beck}\ \emph {et~al.}(2021)\citenamefont {Beck},
  \citenamefont {Lu}, \citenamefont {Sushko}, \citenamefont {Xu},\ and\
  \citenamefont {Li}}]{Beck2021}%
  \BibitemOpen
  \bibfield  {author} {\bibinfo {author} {\bibfnamefont {R.~A.}\ \bibnamefont
  {Beck}}, \bibinfo {author} {\bibfnamefont {L.}~\bibnamefont {Lu}}, \bibinfo
  {author} {\bibfnamefont {P.~V.}\ \bibnamefont {Sushko}}, \bibinfo {author}
  {\bibfnamefont {X.}~\bibnamefont {Xu}},\ and\ \bibinfo {author}
  {\bibfnamefont {X.}~\bibnamefont {Li}},\ }\bibfield  {title} {\bibinfo
  {title} {Defect-induced magnetic skyrmion in a two-dimensional chromium
  triiodide monolayer},\ }\href {https://doi.org/10.1021/jacsau.1c00142}
  {\bibfield  {journal} {\bibinfo  {journal} {JACS Au}\ }\textbf {\bibinfo
  {volume} {1}},\ \bibinfo {pages} {1362} (\bibinfo {year} {2021})}\BibitemShut
  {NoStop}%
\bibitem [{\citenamefont {Fumega}\ and\ \citenamefont
  {Lado}(2023)}]{Fumega_2023}%
  \BibitemOpen
  \bibfield  {author} {\bibinfo {author} {\bibfnamefont {A.~O.}\ \bibnamefont
  {Fumega}}\ and\ \bibinfo {author} {\bibfnamefont {J.~L.}\ \bibnamefont
  {Lado}},\ }\bibfield  {title} {\bibinfo {title} {Moiré-driven multiferroic
  order in twisted {CrCl3, CrBr3} and {CrI3} bilayers},\ }\href
  {https://doi.org/10.1088/2053-1583/acc671} {\bibfield  {journal} {\bibinfo
  {journal} {2D Materials}\ }\textbf {\bibinfo {volume} {10}},\ \bibinfo
  {pages} {025026} (\bibinfo {year} {2023})}\BibitemShut {NoStop}%
\bibitem [{\citenamefont {Xie}\ \emph {et~al.}(2022)\citenamefont {Xie},
  \citenamefont {Luo}, \citenamefont {Ye}, \citenamefont {Ye}, \citenamefont
  {Ge}, \citenamefont {Sung}, \citenamefont {Rennich}, \citenamefont {Yan},
  \citenamefont {Fu}, \citenamefont {Tian}, \citenamefont {Lei}, \citenamefont
  {Hovden}, \citenamefont {Sun}, \citenamefont {He},\ and\ \citenamefont
  {Zhao}}]{Xie2022}%
  \BibitemOpen
  \bibfield  {author} {\bibinfo {author} {\bibfnamefont {H.}~\bibnamefont
  {Xie}}, \bibinfo {author} {\bibfnamefont {X.}~\bibnamefont {Luo}}, \bibinfo
  {author} {\bibfnamefont {G.}~\bibnamefont {Ye}}, \bibinfo {author}
  {\bibfnamefont {Z.}~\bibnamefont {Ye}}, \bibinfo {author} {\bibfnamefont
  {H.}~\bibnamefont {Ge}}, \bibinfo {author} {\bibfnamefont {S.~H.}\
  \bibnamefont {Sung}}, \bibinfo {author} {\bibfnamefont {E.}~\bibnamefont
  {Rennich}}, \bibinfo {author} {\bibfnamefont {S.}~\bibnamefont {Yan}},
  \bibinfo {author} {\bibfnamefont {Y.}~\bibnamefont {Fu}}, \bibinfo {author}
  {\bibfnamefont {S.}~\bibnamefont {Tian}}, \bibinfo {author} {\bibfnamefont
  {H.}~\bibnamefont {Lei}}, \bibinfo {author} {\bibfnamefont {R.}~\bibnamefont
  {Hovden}}, \bibinfo {author} {\bibfnamefont {K.}~\bibnamefont {Sun}},
  \bibinfo {author} {\bibfnamefont {R.}~\bibnamefont {He}},\ and\ \bibinfo
  {author} {\bibfnamefont {L.}~\bibnamefont {Zhao}},\ }\bibfield  {title}
  {\bibinfo {title} {Twist engineering of the two-dimensional magnetism in
  double bilayer chromium triiodide homostructures},\ }\href
  {https://doi.org/10.1038/s41567-021-01408-8} {\bibfield  {journal} {\bibinfo
  {journal} {Nature Physics}\ }\textbf {\bibinfo {volume} {18}},\ \bibinfo
  {pages} {30} (\bibinfo {year} {2022})}\BibitemShut {NoStop}%
\bibitem [{\citenamefont {Dolui}\ \emph {et~al.}(2020)\citenamefont {Dolui},
  \citenamefont {Petrovi{\'{c}}}, \citenamefont {Zollner}, \citenamefont
  {Plech{\'a}{\v{c}}}, \citenamefont {Fabian},\ and\ \citenamefont
  {Nikoli{\'{c}}}}]{Dolui2020}%
  \BibitemOpen
  \bibfield  {author} {\bibinfo {author} {\bibfnamefont {K.}~\bibnamefont
  {Dolui}}, \bibinfo {author} {\bibfnamefont {M.~D.}\ \bibnamefont
  {Petrovi{\'{c}}}}, \bibinfo {author} {\bibfnamefont {K.}~\bibnamefont
  {Zollner}}, \bibinfo {author} {\bibfnamefont {P.}~\bibnamefont
  {Plech{\'a}{\v{c}}}}, \bibinfo {author} {\bibfnamefont {J.}~\bibnamefont
  {Fabian}},\ and\ \bibinfo {author} {\bibfnamefont {B.~K.}\ \bibnamefont
  {Nikoli{\'{c}}}},\ }\bibfield  {title} {\bibinfo {title} {Proximity
  spin--orbit torque on a two-dimensional magnet within van der waals
  heterostructure: Current-driven antiferromagnet-to-ferromagnet reversible
  nonequilibrium phase transition in bilayer {CrI3}},\ }\href
  {https://doi.org/10.1021/acs.nanolett.9b04556} {\bibfield  {journal}
  {\bibinfo  {journal} {Nano Letters}\ }\textbf {\bibinfo {volume} {20}},\
  \bibinfo {pages} {2288} (\bibinfo {year} {2020})}\BibitemShut {NoStop}%
\bibitem [{\citenamefont {Song}\ and\ \citenamefont
  {Fal'ko}(2022)}]{FalkoSong2022}%
  \BibitemOpen
  \bibfield  {author} {\bibinfo {author} {\bibfnamefont {K.~W.}\ \bibnamefont
  {Song}}\ and\ \bibinfo {author} {\bibfnamefont {V.~I.}\ \bibnamefont
  {Fal'ko}},\ }\bibfield  {title} {\bibinfo {title} {Superexchange and
  spin-orbit coupling in monolayer and bilayer chromium trihalides},\ }\href
  {https://doi.org/10.1103/PhysRevB.106.245111} {\bibfield  {journal} {\bibinfo
   {journal} {Phys. Rev. B}\ }\textbf {\bibinfo {volume} {106}},\ \bibinfo
  {pages} {245111} (\bibinfo {year} {2022})}\BibitemShut {NoStop}%
\bibitem [{\citenamefont {Song}(2023)}]{Song2023}%
  \BibitemOpen
  \bibfield  {author} {\bibinfo {author} {\bibfnamefont {K.~W.}\ \bibnamefont
  {Song}},\ }\bibfield  {title} {\bibinfo {title} {Interlayer superexchange in
  bilayer chromium trihalides},\ }\href
  {https://doi.org/10.1103/PhysRevB.107.245133} {\bibfield  {journal} {\bibinfo
   {journal} {Phys. Rev. B}\ }\textbf {\bibinfo {volume} {107}},\ \bibinfo
  {pages} {245133} (\bibinfo {year} {2023})}\BibitemShut {NoStop}%
\bibitem [{\citenamefont {Soriano}\ \emph {et~al.}(2019)\citenamefont
  {Soriano}, \citenamefont {Cardoso},\ and\ \citenamefont
  {Fernández-Rossier}}]{SORIANO2019}%
  \BibitemOpen
  \bibfield  {author} {\bibinfo {author} {\bibfnamefont {D.}~\bibnamefont
  {Soriano}}, \bibinfo {author} {\bibfnamefont {C.}~\bibnamefont {Cardoso}},\
  and\ \bibinfo {author} {\bibfnamefont {J.}~\bibnamefont
  {Fernández-Rossier}},\ }\bibfield  {title} {\bibinfo {title} {Interplay
  between interlayer exchange and stacking in {CrI3} bilayers},\ }\href
  {https://doi.org/https://doi.org/10.1016/j.ssc.2019.113662} {\bibfield
  {journal} {\bibinfo  {journal} {Solid State Communications}\ }\textbf
  {\bibinfo {volume} {299}},\ \bibinfo {pages} {113662} (\bibinfo {year}
  {2019})}\BibitemShut {NoStop}%
\bibitem [{\citenamefont {Beck}\ \emph {et~al.}(2022)\citenamefont {Beck},
  \citenamefont {Sun}, \citenamefont {Xu}, \citenamefont {Gamelin},
  \citenamefont {Cao},\ and\ \citenamefont {Li}}]{Beck2022}%
  \BibitemOpen
  \bibfield  {author} {\bibinfo {author} {\bibfnamefont {R.~A.}\ \bibnamefont
  {Beck}}, \bibinfo {author} {\bibfnamefont {S.}~\bibnamefont {Sun}}, \bibinfo
  {author} {\bibfnamefont {X.}~\bibnamefont {Xu}}, \bibinfo {author}
  {\bibfnamefont {D.~R.}\ \bibnamefont {Gamelin}}, \bibinfo {author}
  {\bibfnamefont {T.}~\bibnamefont {Cao}},\ and\ \bibinfo {author}
  {\bibfnamefont {X.}~\bibnamefont {Li}},\ }\bibfield  {title} {\bibinfo
  {title} {Understanding external pressure effects and interlayer orbital
  exchange pathways in the two-dimensional magnet-chromium triiodide},\ }\href
  {https://doi.org/10.1021/acs.jpcc.2c03884} {\bibfield  {journal} {\bibinfo
  {journal} {The Journal of Physical Chemistry C}\ }\textbf {\bibinfo {volume}
  {126}},\ \bibinfo {pages} {19327} (\bibinfo {year} {2022})}\BibitemShut
  {NoStop}%
\bibitem [{\citenamefont {Ghosh}\ \emph {et~al.}(2022)\citenamefont {Ghosh},
  \citenamefont {Singh}, \citenamefont {Aramaki}, \citenamefont {Mu},
  \citenamefont {Borisov}, \citenamefont {Kvashnin}, \citenamefont {Haider},
  \citenamefont {Jonak}, \citenamefont {Chareev}, \citenamefont {Medvedev},
  \citenamefont {Klingeler}, \citenamefont {Mito}, \citenamefont
  {Abdul-Hafidh}, \citenamefont {Vejpravova}, \citenamefont
  {Kalb\`a\ifmmode~\check{c}\else \v{c}\fi{}}, \citenamefont {Ahuja},
  \citenamefont {Eriksson},\ and\ \citenamefont
  {Abde-Hafiez}}]{PhysRevB.105.L081104}%
  \BibitemOpen
  \bibfield  {author} {\bibinfo {author} {\bibfnamefont {A.}~\bibnamefont
  {Ghosh}}, \bibinfo {author} {\bibfnamefont {D.}~\bibnamefont {Singh}},
  \bibinfo {author} {\bibfnamefont {T.}~\bibnamefont {Aramaki}}, \bibinfo
  {author} {\bibfnamefont {Q.}~\bibnamefont {Mu}}, \bibinfo {author}
  {\bibfnamefont {V.}~\bibnamefont {Borisov}}, \bibinfo {author} {\bibfnamefont
  {Y.}~\bibnamefont {Kvashnin}}, \bibinfo {author} {\bibfnamefont
  {G.}~\bibnamefont {Haider}}, \bibinfo {author} {\bibfnamefont
  {M.}~\bibnamefont {Jonak}}, \bibinfo {author} {\bibfnamefont
  {D.}~\bibnamefont {Chareev}}, \bibinfo {author} {\bibfnamefont {S.~A.}\
  \bibnamefont {Medvedev}}, \bibinfo {author} {\bibfnamefont {R.}~\bibnamefont
  {Klingeler}}, \bibinfo {author} {\bibfnamefont {M.}~\bibnamefont {Mito}},
  \bibinfo {author} {\bibfnamefont {E.~H.}\ \bibnamefont {Abdul-Hafidh}},
  \bibinfo {author} {\bibfnamefont {J.}~\bibnamefont {Vejpravova}}, \bibinfo
  {author} {\bibfnamefont {M.}~\bibnamefont {Kalb\`a\ifmmode~\check{c}\else
  \v{c}\fi{}}}, \bibinfo {author} {\bibfnamefont {R.}~\bibnamefont {Ahuja}},
  \bibinfo {author} {\bibfnamefont {O.}~\bibnamefont {Eriksson}},\ and\
  \bibinfo {author} {\bibfnamefont {M.}~\bibnamefont {Abde-Hafiez}},\
  }\bibfield  {title} {\bibinfo {title} {Exotic magnetic and electronic
  properties of layered ${\mathrm{cri}}_{3}$ single crystals under high
  pressure},\ }\href {https://doi.org/10.1103/PhysRevB.105.L081104} {\bibfield
  {journal} {\bibinfo  {journal} {Phys. Rev. B}\ }\textbf {\bibinfo {volume}
  {105}},\ \bibinfo {pages} {L081104} (\bibinfo {year} {2022})}\BibitemShut
  {NoStop}%
\bibitem [{\citenamefont {Klein}\ \emph {et~al.}(2019)\citenamefont {Klein},
  \citenamefont {MacNeill}, \citenamefont {Song}, \citenamefont {Larson},
  \citenamefont {Fang}, \citenamefont {Xu}, \citenamefont {Ribeiro},
  \citenamefont {Canfield}, \citenamefont {Kaxiras}, \citenamefont {Comin},\
  and\ \citenamefont {Jarillo-Herrero}}]{Klein2019}%
  \BibitemOpen
  \bibfield  {author} {\bibinfo {author} {\bibfnamefont {D.~R.}\ \bibnamefont
  {Klein}}, \bibinfo {author} {\bibfnamefont {D.}~\bibnamefont {MacNeill}},
  \bibinfo {author} {\bibfnamefont {Q.}~\bibnamefont {Song}}, \bibinfo {author}
  {\bibfnamefont {D.~T.}\ \bibnamefont {Larson}}, \bibinfo {author}
  {\bibfnamefont {S.}~\bibnamefont {Fang}}, \bibinfo {author} {\bibfnamefont
  {M.}~\bibnamefont {Xu}}, \bibinfo {author} {\bibfnamefont {R.~A.}\
  \bibnamefont {Ribeiro}}, \bibinfo {author} {\bibfnamefont {P.~C.}\
  \bibnamefont {Canfield}}, \bibinfo {author} {\bibfnamefont {E.}~\bibnamefont
  {Kaxiras}}, \bibinfo {author} {\bibfnamefont {R.}~\bibnamefont {Comin}},\
  and\ \bibinfo {author} {\bibfnamefont {P.}~\bibnamefont {Jarillo-Herrero}},\
  }\bibfield  {title} {\bibinfo {title} {Enhancement of interlayer exchange in
  an ultrathin two-dimensional magnet},\ }\href
  {https://doi.org/10.1038/s41567-019-0651-0} {\bibfield  {journal} {\bibinfo
  {journal} {Nature Physics}\ }\textbf {\bibinfo {volume} {15}},\ \bibinfo
  {pages} {1255} (\bibinfo {year} {2019})}\BibitemShut {NoStop}%
\bibitem [{\citenamefont {Webster}\ and\ \citenamefont
  {Yan}(2018)}]{PhysRevB.98.144411}%
  \BibitemOpen
  \bibfield  {author} {\bibinfo {author} {\bibfnamefont {L.}~\bibnamefont
  {Webster}}\ and\ \bibinfo {author} {\bibfnamefont {J.-A.}\ \bibnamefont
  {Yan}},\ }\bibfield  {title} {\bibinfo {title} {Strain-tunable magnetic
  anisotropy in monolayer ${\mathrm{crcl}}_{3}$, ${\mathrm{crbr}}_{3}$, and
  ${\mathrm{cri}}_{3}$},\ }\href {https://doi.org/10.1103/PhysRevB.98.144411}
  {\bibfield  {journal} {\bibinfo  {journal} {Phys. Rev. B}\ }\textbf {\bibinfo
  {volume} {98}},\ \bibinfo {pages} {144411} (\bibinfo {year}
  {2018})}\BibitemShut {NoStop}%
\bibitem [{\citenamefont {Zhong}\ \emph {et~al.}(2020)\citenamefont {Zhong},
  \citenamefont {Seyler}, \citenamefont {Linpeng}, \citenamefont {Wilson},
  \citenamefont {Taniguchi}, \citenamefont {Watanabe}, \citenamefont {McGuire},
  \citenamefont {Fu}, \citenamefont {Xiao}, \citenamefont {Yao},\ and\
  \citenamefont {Xu}}]{Zhong2020}%
  \BibitemOpen
  \bibfield  {author} {\bibinfo {author} {\bibfnamefont {D.}~\bibnamefont
  {Zhong}}, \bibinfo {author} {\bibfnamefont {K.~L.}\ \bibnamefont {Seyler}},
  \bibinfo {author} {\bibfnamefont {X.}~\bibnamefont {Linpeng}}, \bibinfo
  {author} {\bibfnamefont {N.~P.}\ \bibnamefont {Wilson}}, \bibinfo {author}
  {\bibfnamefont {T.}~\bibnamefont {Taniguchi}}, \bibinfo {author}
  {\bibfnamefont {K.}~\bibnamefont {Watanabe}}, \bibinfo {author}
  {\bibfnamefont {M.~A.}\ \bibnamefont {McGuire}}, \bibinfo {author}
  {\bibfnamefont {K.-M.~C.}\ \bibnamefont {Fu}}, \bibinfo {author}
  {\bibfnamefont {D.}~\bibnamefont {Xiao}}, \bibinfo {author} {\bibfnamefont
  {W.}~\bibnamefont {Yao}},\ and\ \bibinfo {author} {\bibfnamefont
  {X.}~\bibnamefont {Xu}},\ }\bibfield  {title} {\bibinfo {title}
  {Layer-resolved magnetic proximity effect in van der waals
  heterostructures},\ }\href {https://doi.org/10.1038/s41565-019-0629-1}
  {\bibfield  {journal} {\bibinfo  {journal} {Nature Nanotechnology}\ }\textbf
  {\bibinfo {volume} {15}},\ \bibinfo {pages} {187} (\bibinfo {year}
  {2020})}\BibitemShut {NoStop}%
\bibitem [{\citenamefont {Wu}\ \emph {et~al.}(2021)\citenamefont {Wu},
  \citenamefont {Cui}, \citenamefont {Zhu}, \citenamefont {Liu}, \citenamefont
  {Wang}, \citenamefont {Zhang}, \citenamefont {Zheng}, \citenamefont {Shen},
  \citenamefont {Cui}, \citenamefont {Yang},\ and\ \citenamefont
  {Wang}}]{Wu2021}%
  \BibitemOpen
  \bibfield  {author} {\bibinfo {author} {\bibfnamefont {Y.}~\bibnamefont
  {Wu}}, \bibinfo {author} {\bibfnamefont {Q.}~\bibnamefont {Cui}}, \bibinfo
  {author} {\bibfnamefont {M.}~\bibnamefont {Zhu}}, \bibinfo {author}
  {\bibfnamefont {X.}~\bibnamefont {Liu}}, \bibinfo {author} {\bibfnamefont
  {Y.}~\bibnamefont {Wang}}, \bibinfo {author} {\bibfnamefont {J.}~\bibnamefont
  {Zhang}}, \bibinfo {author} {\bibfnamefont {X.}~\bibnamefont {Zheng}},
  \bibinfo {author} {\bibfnamefont {J.}~\bibnamefont {Shen}}, \bibinfo {author}
  {\bibfnamefont {P.}~\bibnamefont {Cui}}, \bibinfo {author} {\bibfnamefont
  {H.}~\bibnamefont {Yang}},\ and\ \bibinfo {author} {\bibfnamefont
  {S.}~\bibnamefont {Wang}},\ }\bibfield  {title} {\bibinfo {title} {Magnetic
  exchange field modulation of quantum hall ferromagnetism in 2d van der waals
  $\rm crcl3/graphene$ heterostructures},\ }\href
  {https://doi.org/10.1021/acsami.1c00551} {\bibfield  {journal} {\bibinfo
  {journal} {ACS Applied Materials {\&} Interfaces}\ }\textbf {\bibinfo
  {volume} {13}},\ \bibinfo {pages} {10656} (\bibinfo {year}
  {2021})}\BibitemShut {NoStop}%
\bibitem [{\citenamefont {Wu}\ \emph {et~al.}(2019{\natexlab{b}})\citenamefont
  {Wu}, \citenamefont {Yu},\ and\ \citenamefont {Yuan}}]{Wu2019}%
  \BibitemOpen
  \bibfield  {author} {\bibinfo {author} {\bibfnamefont {Z.}~\bibnamefont
  {Wu}}, \bibinfo {author} {\bibfnamefont {J.}~\bibnamefont {Yu}},\ and\
  \bibinfo {author} {\bibfnamefont {S.}~\bibnamefont {Yuan}},\ }\bibfield
  {title} {\bibinfo {title} {Strain-tunable magnetic and electronic properties
  of monolayer cri3},\ }\href {https://doi.org/10.1039/C8CP07067A} {\bibfield
  {journal} {\bibinfo  {journal} {Physical Chemistry Chemical Physics}\
  }\textbf {\bibinfo {volume} {21}},\ \bibinfo {pages} {7750} (\bibinfo {year}
  {2019}{\natexlab{b}})}\BibitemShut {NoStop}%
\bibitem [{\citenamefont {Georgescu}\ \emph {et~al.}(2022)\citenamefont
  {Georgescu}, \citenamefont {Millis},\ and\ \citenamefont
  {Rondinelli}}]{Georgescu2022}%
  \BibitemOpen
  \bibfield  {author} {\bibinfo {author} {\bibfnamefont {A.~B.}\ \bibnamefont
  {Georgescu}}, \bibinfo {author} {\bibfnamefont {A.~J.}\ \bibnamefont
  {Millis}},\ and\ \bibinfo {author} {\bibfnamefont {J.~M.}\ \bibnamefont
  {Rondinelli}},\ }\bibfield  {title} {\bibinfo {title} {Trigonal symmetry
  breaking and its electronic effects in the two-dimensional dihalides
  $m{X}_{2}$ and trihalides $m{X}_{3}$},\ }\href
  {https://doi.org/10.1103/PhysRevB.105.245153} {\bibfield  {journal} {\bibinfo
   {journal} {Phys. Rev. B}\ }\textbf {\bibinfo {volume} {105}},\ \bibinfo
  {pages} {245153} (\bibinfo {year} {2022})}\BibitemShut {NoStop}%
\bibitem [{\citenamefont {Grzeszczyk}\ \emph {et~al.}(2023)\citenamefont
  {Grzeszczyk}, \citenamefont {Acharya}, \citenamefont {Pashov}, \citenamefont
  {Chen}, \citenamefont {Vaklinova}, \citenamefont {van Schilfgaarde},
  \citenamefont {Watanabe}, \citenamefont {Taniguchi}, \citenamefont
  {Novoselov}, \citenamefont {Katsnelson},\ and\ \citenamefont
  {Koperski}}]{https://doi.org/10.1002/adma.202209513}%
  \BibitemOpen
  \bibfield  {author} {\bibinfo {author} {\bibfnamefont {M.}~\bibnamefont
  {Grzeszczyk}}, \bibinfo {author} {\bibfnamefont {S.}~\bibnamefont {Acharya}},
  \bibinfo {author} {\bibfnamefont {D.}~\bibnamefont {Pashov}}, \bibinfo
  {author} {\bibfnamefont {Z.}~\bibnamefont {Chen}}, \bibinfo {author}
  {\bibfnamefont {K.}~\bibnamefont {Vaklinova}}, \bibinfo {author}
  {\bibfnamefont {M.}~\bibnamefont {van Schilfgaarde}}, \bibinfo {author}
  {\bibfnamefont {K.}~\bibnamefont {Watanabe}}, \bibinfo {author}
  {\bibfnamefont {T.}~\bibnamefont {Taniguchi}}, \bibinfo {author}
  {\bibfnamefont {K.~S.}\ \bibnamefont {Novoselov}}, \bibinfo {author}
  {\bibfnamefont {M.~I.}\ \bibnamefont {Katsnelson}},\ and\ \bibinfo {author}
  {\bibfnamefont {M.}~\bibnamefont {Koperski}},\ }\bibfield  {title} {\bibinfo
  {title} {Strongly correlated exciton-magnetization system for optical spin
  pumping in crbr3 and cri3.},\ }\href
  {https://doi.org/https://doi.org/10.1002/adma.202209513} {\bibfield
  {journal} {\bibinfo  {journal} {Advanced Materials}\ }\textbf {\bibinfo
  {volume} {35}},\ \bibinfo {pages} {2209513} (\bibinfo {year} {2023})},\
  \Eprint
  {https://arxiv.org/abs/https://onlinelibrary.wiley.com/doi/pdf/10.1002/adma.202209513}
  {https://onlinelibrary.wiley.com/doi/pdf/10.1002/adma.202209513} \BibitemShut
  {NoStop}%
\bibitem [{\citenamefont {Wu}\ \emph {et~al.}(2019{\natexlab{c}})\citenamefont
  {Wu}, \citenamefont {Yu},\ and\ \citenamefont {Yuan}}]{C8CP07067A}%
  \BibitemOpen
  \bibfield  {author} {\bibinfo {author} {\bibfnamefont {Z.}~\bibnamefont
  {Wu}}, \bibinfo {author} {\bibfnamefont {J.}~\bibnamefont {Yu}},\ and\
  \bibinfo {author} {\bibfnamefont {S.}~\bibnamefont {Yuan}},\ }\bibfield
  {title} {\bibinfo {title} {Strain-tunable magnetic and electronic properties
  of monolayer cri3},\ }\href {https://doi.org/10.1039/C8CP07067A} {\bibfield
  {journal} {\bibinfo  {journal} {Phys. Chem. Chem. Phys.}\ }\textbf {\bibinfo
  {volume} {21}},\ \bibinfo {pages} {7750} (\bibinfo {year}
  {2019}{\natexlab{c}})}\BibitemShut {NoStop}%
\bibitem [{\citenamefont {Zhang}\ \emph {et~al.}(2018)\citenamefont {Zhang},
  \citenamefont {Zhao}, \citenamefont {Zhou}, \citenamefont {Xue},
  \citenamefont {Ma},\ and\ \citenamefont {Yang}}]{Zhang2018}%
  \BibitemOpen
  \bibfield  {author} {\bibinfo {author} {\bibfnamefont {J.}~\bibnamefont
  {Zhang}}, \bibinfo {author} {\bibfnamefont {B.}~\bibnamefont {Zhao}},
  \bibinfo {author} {\bibfnamefont {T.}~\bibnamefont {Zhou}}, \bibinfo {author}
  {\bibfnamefont {Y.}~\bibnamefont {Xue}}, \bibinfo {author} {\bibfnamefont
  {C.}~\bibnamefont {Ma}},\ and\ \bibinfo {author} {\bibfnamefont
  {Z.}~\bibnamefont {Yang}},\ }\bibfield  {title} {\bibinfo {title} {Strong
  magnetization and chern insulators in compressed
  ${\mathrm{graphene}\text{/}\mathrm{cri}}_{3}$ van der waals
  heterostructures},\ }\href {https://doi.org/10.1103/PhysRevB.97.085401}
  {\bibfield  {journal} {\bibinfo  {journal} {Phys. Rev. B}\ }\textbf {\bibinfo
  {volume} {97}},\ \bibinfo {pages} {085401} (\bibinfo {year}
  {2018})}\BibitemShut {NoStop}%
\bibitem [{\citenamefont {Ubrig}\ \emph {et~al.}(2019)\citenamefont {Ubrig},
  \citenamefont {Wang}, \citenamefont {Teyssier}, \citenamefont {Taniguchi},
  \citenamefont {Watanabe}, \citenamefont {Giannini}, \citenamefont
  {Morpurgo},\ and\ \citenamefont {Gibertini}}]{Ubrig_2020}%
  \BibitemOpen
  \bibfield  {author} {\bibinfo {author} {\bibfnamefont {N.}~\bibnamefont
  {Ubrig}}, \bibinfo {author} {\bibfnamefont {Z.}~\bibnamefont {Wang}},
  \bibinfo {author} {\bibfnamefont {J.}~\bibnamefont {Teyssier}}, \bibinfo
  {author} {\bibfnamefont {T.}~\bibnamefont {Taniguchi}}, \bibinfo {author}
  {\bibfnamefont {K.}~\bibnamefont {Watanabe}}, \bibinfo {author}
  {\bibfnamefont {E.}~\bibnamefont {Giannini}}, \bibinfo {author}
  {\bibfnamefont {A.~F.}\ \bibnamefont {Morpurgo}},\ and\ \bibinfo {author}
  {\bibfnamefont {M.}~\bibnamefont {Gibertini}},\ }\bibfield  {title} {\bibinfo
  {title} {Low-temperature monoclinic layer stacking in atomically thin cri3
  crystals},\ }\href {https://doi.org/10.1088/2053-1583/ab4c64} {\bibfield
  {journal} {\bibinfo  {journal} {2D Materials}\ }\textbf {\bibinfo {volume}
  {7}},\ \bibinfo {pages} {015007} (\bibinfo {year} {2019})}\BibitemShut
  {NoStop}%
\bibitem [{\citenamefont {Liu}\ \emph {et~al.}(2022)\citenamefont {Liu},
  \citenamefont {Guo}, \citenamefont {Chen}, \citenamefont {Gong},
  \citenamefont {Li}, \citenamefont {Niu}, \citenamefont {Cheng}, \citenamefont
  {Lu}, \citenamefont {Deng},\ and\ \citenamefont
  {Peng}}]{LiuGuoChenGongLiNiuChengLuDengPeng+2022+4409+4417}%
  \BibitemOpen
  \bibfield  {author} {\bibinfo {author} {\bibfnamefont {Z.}~\bibnamefont
  {Liu}}, \bibinfo {author} {\bibfnamefont {Y.}~\bibnamefont {Guo}}, \bibinfo
  {author} {\bibfnamefont {Z.}~\bibnamefont {Chen}}, \bibinfo {author}
  {\bibfnamefont {T.}~\bibnamefont {Gong}}, \bibinfo {author} {\bibfnamefont
  {Y.}~\bibnamefont {Li}}, \bibinfo {author} {\bibfnamefont {Y.}~\bibnamefont
  {Niu}}, \bibinfo {author} {\bibfnamefont {Y.}~\bibnamefont {Cheng}}, \bibinfo
  {author} {\bibfnamefont {H.}~\bibnamefont {Lu}}, \bibinfo {author}
  {\bibfnamefont {L.}~\bibnamefont {Deng}},\ and\ \bibinfo {author}
  {\bibfnamefont {B.}~\bibnamefont {Peng}},\ }\bibfield  {title} {\bibinfo
  {title} {Observation of intrinsic crystal phase in bare few-layer cri3},\
  }\href {https://doi.org/doi:10.1515/nanoph-2022-0246} {\bibfield  {journal}
  {\bibinfo  {journal} {Nanophotonics}\ }\textbf {\bibinfo {volume} {11}},\
  \bibinfo {pages} {4409} (\bibinfo {year} {2022})}\BibitemShut {NoStop}%
\bibitem [{\citenamefont {Li}\ \emph {et~al.}(2019)\citenamefont {Li},
  \citenamefont {Jiang}, \citenamefont {Sivadas}, \citenamefont {Wang},
  \citenamefont {Xu}, \citenamefont {Weber}, \citenamefont {Goldberger},
  \citenamefont {Watanabe}, \citenamefont {Taniguchi}, \citenamefont {Fennie},
  \citenamefont {Fai~Mak},\ and\ \citenamefont {Shan}}]{Li2019}%
  \BibitemOpen
  \bibfield  {author} {\bibinfo {author} {\bibfnamefont {T.}~\bibnamefont
  {Li}}, \bibinfo {author} {\bibfnamefont {S.}~\bibnamefont {Jiang}}, \bibinfo
  {author} {\bibfnamefont {N.}~\bibnamefont {Sivadas}}, \bibinfo {author}
  {\bibfnamefont {Z.}~\bibnamefont {Wang}}, \bibinfo {author} {\bibfnamefont
  {Y.}~\bibnamefont {Xu}}, \bibinfo {author} {\bibfnamefont {D.}~\bibnamefont
  {Weber}}, \bibinfo {author} {\bibfnamefont {J.~E.}\ \bibnamefont
  {Goldberger}}, \bibinfo {author} {\bibfnamefont {K.}~\bibnamefont
  {Watanabe}}, \bibinfo {author} {\bibfnamefont {T.}~\bibnamefont {Taniguchi}},
  \bibinfo {author} {\bibfnamefont {C.~J.}\ \bibnamefont {Fennie}}, \bibinfo
  {author} {\bibfnamefont {K.}~\bibnamefont {Fai~Mak}},\ and\ \bibinfo {author}
  {\bibfnamefont {J.}~\bibnamefont {Shan}},\ }\bibfield  {title} {\bibinfo
  {title} {Pressure-controlled interlayer magnetism in atomically thin
  {CrI3}},\ }\href {https://doi.org/10.1038/s41563-019-0506-1} {\bibfield
  {journal} {\bibinfo  {journal} {Nature Materials}\ }\textbf {\bibinfo
  {volume} {18}},\ \bibinfo {pages} {1303} (\bibinfo {year}
  {2019})}\BibitemShut {NoStop}%
\bibitem [{\citenamefont {Djurdji\ifmmode \acute{c}\else~\'{c}\fi{} Mijin}\
  \emph {et~al.}(2018)\citenamefont {Djurdji\ifmmode \acute{c}\else~\'{c}\fi{}
  Mijin}, \citenamefont {\ifmmode \check{S}\else
  \v{S}\fi{}olaji\ifmmode~\acute{c}\else \'{c}\fi{}}, \citenamefont {Pe\ifmmode
  \check{s}\else \v{s}\fi{}i\ifmmode~\acute{c}\else \'{c}\fi{}}, \citenamefont
  {\ifmmode \check{S}\else \v{S}\fi{}\ifmmode \acute{c}\else
  \'{c}\fi{}epanovi\ifmmode~\acute{c}\else \'{c}\fi{}}, \citenamefont {Liu},
  \citenamefont {Baum}, \citenamefont {Petrovic}, \citenamefont
  {Lazarevi\ifmmode~\acute{c}\else \'{c}\fi{}},\ and\ \citenamefont
  {Popovi\ifmmode~\acute{c}\else \'{c}\fi{}}}]{PhysRevB.98.104307}%
  \BibitemOpen
  \bibfield  {author} {\bibinfo {author} {\bibfnamefont {S.}~\bibnamefont
  {Djurdji\ifmmode \acute{c}\else~\'{c}\fi{} Mijin}}, \bibinfo {author}
  {\bibfnamefont {A.}~\bibnamefont {\ifmmode \check{S}\else
  \v{S}\fi{}olaji\ifmmode~\acute{c}\else \'{c}\fi{}}}, \bibinfo {author}
  {\bibfnamefont {J.}~\bibnamefont {Pe\ifmmode \check{s}\else
  \v{s}\fi{}i\ifmmode~\acute{c}\else \'{c}\fi{}}}, \bibinfo {author}
  {\bibfnamefont {M.}~\bibnamefont {\ifmmode \check{S}\else \v{S}\fi{}\ifmmode
  \acute{c}\else \'{c}\fi{}epanovi\ifmmode~\acute{c}\else \'{c}\fi{}}},
  \bibinfo {author} {\bibfnamefont {Y.}~\bibnamefont {Liu}}, \bibinfo {author}
  {\bibfnamefont {A.}~\bibnamefont {Baum}}, \bibinfo {author} {\bibfnamefont
  {C.}~\bibnamefont {Petrovic}}, \bibinfo {author} {\bibfnamefont
  {N.}~\bibnamefont {Lazarevi\ifmmode~\acute{c}\else \'{c}\fi{}}},\ and\
  \bibinfo {author} {\bibfnamefont {Z.~V.}\ \bibnamefont
  {Popovi\ifmmode~\acute{c}\else \'{c}\fi{}}},\ }\bibfield  {title} {\bibinfo
  {title} {Lattice dynamics and phase transition in ${\mathrm{cri}}_{3}$ single
  crystals},\ }\href {https://doi.org/10.1103/PhysRevB.98.104307} {\bibfield
  {journal} {\bibinfo  {journal} {Phys. Rev. B}\ }\textbf {\bibinfo {volume}
  {98}},\ \bibinfo {pages} {104307} (\bibinfo {year} {2018})}\BibitemShut
  {NoStop}%
\bibitem [{\citenamefont {Giannozzi}\ \emph {et~al.}(2009)\citenamefont
  {Giannozzi}, \citenamefont {Baroni}, \citenamefont {Bonini}, \citenamefont
  {Calandra}, \citenamefont {Car}, \citenamefont {Cavazzoni}, \citenamefont
  {Ceresoli}, \citenamefont {Chiarotti}, \citenamefont {Cococcioni},
  \citenamefont {Dabo}, \citenamefont {Corso}, \citenamefont {de~Gironcoli},
  \citenamefont {Fabris}, \citenamefont {Fratesi}, \citenamefont {Gebauer},
  \citenamefont {Gerstmann}, \citenamefont {Gougoussis}, \citenamefont
  {Kokalj}, \citenamefont {Lazzeri}, \citenamefont {Martin-Samos},
  \citenamefont {Marzari}, \citenamefont {Mauri}, \citenamefont {Mazzarello},
  \citenamefont {Paolini}, \citenamefont {Pasquarello}, \citenamefont
  {Paulatto}, \citenamefont {Sbraccia}, \citenamefont {Scandolo}, \citenamefont
  {Sclauzero}, \citenamefont {Seitsonen}, \citenamefont {Smogunov},
  \citenamefont {Umari},\ and\ \citenamefont {Wentzcovitch}}]{Giannozzi_2009}%
  \BibitemOpen
  \bibfield  {author} {\bibinfo {author} {\bibfnamefont {P.}~\bibnamefont
  {Giannozzi}}, \bibinfo {author} {\bibfnamefont {S.}~\bibnamefont {Baroni}},
  \bibinfo {author} {\bibfnamefont {N.}~\bibnamefont {Bonini}}, \bibinfo
  {author} {\bibfnamefont {M.}~\bibnamefont {Calandra}}, \bibinfo {author}
  {\bibfnamefont {R.}~\bibnamefont {Car}}, \bibinfo {author} {\bibfnamefont
  {C.}~\bibnamefont {Cavazzoni}}, \bibinfo {author} {\bibfnamefont
  {D.}~\bibnamefont {Ceresoli}}, \bibinfo {author} {\bibfnamefont {G.~L.}\
  \bibnamefont {Chiarotti}}, \bibinfo {author} {\bibfnamefont {M.}~\bibnamefont
  {Cococcioni}}, \bibinfo {author} {\bibfnamefont {I.}~\bibnamefont {Dabo}},
  \bibinfo {author} {\bibfnamefont {A.~D.}\ \bibnamefont {Corso}}, \bibinfo
  {author} {\bibfnamefont {S.}~\bibnamefont {de~Gironcoli}}, \bibinfo {author}
  {\bibfnamefont {S.}~\bibnamefont {Fabris}}, \bibinfo {author} {\bibfnamefont
  {G.}~\bibnamefont {Fratesi}}, \bibinfo {author} {\bibfnamefont
  {R.}~\bibnamefont {Gebauer}}, \bibinfo {author} {\bibfnamefont
  {U.}~\bibnamefont {Gerstmann}}, \bibinfo {author} {\bibfnamefont
  {C.}~\bibnamefont {Gougoussis}}, \bibinfo {author} {\bibfnamefont
  {A.}~\bibnamefont {Kokalj}}, \bibinfo {author} {\bibfnamefont
  {M.}~\bibnamefont {Lazzeri}}, \bibinfo {author} {\bibfnamefont
  {L.}~\bibnamefont {Martin-Samos}}, \bibinfo {author} {\bibfnamefont
  {N.}~\bibnamefont {Marzari}}, \bibinfo {author} {\bibfnamefont
  {F.}~\bibnamefont {Mauri}}, \bibinfo {author} {\bibfnamefont
  {R.}~\bibnamefont {Mazzarello}}, \bibinfo {author} {\bibfnamefont
  {S.}~\bibnamefont {Paolini}}, \bibinfo {author} {\bibfnamefont
  {A.}~\bibnamefont {Pasquarello}}, \bibinfo {author} {\bibfnamefont
  {L.}~\bibnamefont {Paulatto}}, \bibinfo {author} {\bibfnamefont
  {C.}~\bibnamefont {Sbraccia}}, \bibinfo {author} {\bibfnamefont
  {S.}~\bibnamefont {Scandolo}}, \bibinfo {author} {\bibfnamefont
  {G.}~\bibnamefont {Sclauzero}}, \bibinfo {author} {\bibfnamefont {A.~P.}\
  \bibnamefont {Seitsonen}}, \bibinfo {author} {\bibfnamefont {A.}~\bibnamefont
  {Smogunov}}, \bibinfo {author} {\bibfnamefont {P.}~\bibnamefont {Umari}},\
  and\ \bibinfo {author} {\bibfnamefont {R.~M.}\ \bibnamefont {Wentzcovitch}},\
  }\bibfield  {title} {\bibinfo {title} {Quantum espresso: a modular and
  open-source software project for quantum simulations of materials},\ }\href
  {https://doi.org/10.1088/0953-8984/21/39/395502} {\bibfield  {journal}
  {\bibinfo  {journal} {Journal of Physics: Condensed Matter}\ }\textbf
  {\bibinfo {volume} {21}},\ \bibinfo {pages} {395502} (\bibinfo {year}
  {2009})}\BibitemShut {NoStop}%
\bibitem [{\citenamefont {Giannozzi}\ \emph {et~al.}(2017)\citenamefont
  {Giannozzi}, \citenamefont {Andreussi}, \citenamefont {Brumme}, \citenamefont
  {Bunau}, \citenamefont {Nardelli}, \citenamefont {Calandra}, \citenamefont
  {Car}, \citenamefont {Cavazzoni}, \citenamefont {Ceresoli}, \citenamefont
  {Cococcioni}, \citenamefont {Colonna}, \citenamefont {Carnimeo},
  \citenamefont {Corso}, \citenamefont {de~Gironcoli}, \citenamefont {Delugas},
  \citenamefont {DiStasio}, \citenamefont {Ferretti}, \citenamefont {Floris},
  \citenamefont {Fratesi}, \citenamefont {Fugallo}, \citenamefont {Gebauer},
  \citenamefont {Gerstmann}, \citenamefont {Giustino}, \citenamefont {Gorni},
  \citenamefont {Jia}, \citenamefont {Kawamura}, \citenamefont {Ko},
  \citenamefont {Kokalj}, \citenamefont {Küçükbenli}, \citenamefont
  {Lazzeri}, \citenamefont {Marsili}, \citenamefont {Marzari}, \citenamefont
  {Mauri}, \citenamefont {Nguyen}, \citenamefont {Nguyen}, \citenamefont {de-la
  Roza}, \citenamefont {Paulatto}, \citenamefont {Poncé}, \citenamefont
  {Rocca}, \citenamefont {Sabatini}, \citenamefont {Santra}, \citenamefont
  {Schlipf}, \citenamefont {Seitsonen}, \citenamefont {Smogunov}, \citenamefont
  {Timrov}, \citenamefont {Thonhauser}, \citenamefont {Umari}, \citenamefont
  {Vast}, \citenamefont {Wu},\ and\ \citenamefont {Baroni}}]{Giannozzi_2017}%
  \BibitemOpen
  \bibfield  {author} {\bibinfo {author} {\bibfnamefont {P.}~\bibnamefont
  {Giannozzi}}, \bibinfo {author} {\bibfnamefont {O.}~\bibnamefont
  {Andreussi}}, \bibinfo {author} {\bibfnamefont {T.}~\bibnamefont {Brumme}},
  \bibinfo {author} {\bibfnamefont {O.}~\bibnamefont {Bunau}}, \bibinfo
  {author} {\bibfnamefont {M.~B.}\ \bibnamefont {Nardelli}}, \bibinfo {author}
  {\bibfnamefont {M.}~\bibnamefont {Calandra}}, \bibinfo {author}
  {\bibfnamefont {R.}~\bibnamefont {Car}}, \bibinfo {author} {\bibfnamefont
  {C.}~\bibnamefont {Cavazzoni}}, \bibinfo {author} {\bibfnamefont
  {D.}~\bibnamefont {Ceresoli}}, \bibinfo {author} {\bibfnamefont
  {M.}~\bibnamefont {Cococcioni}}, \bibinfo {author} {\bibfnamefont
  {N.}~\bibnamefont {Colonna}}, \bibinfo {author} {\bibfnamefont
  {I.}~\bibnamefont {Carnimeo}}, \bibinfo {author} {\bibfnamefont {A.~D.}\
  \bibnamefont {Corso}}, \bibinfo {author} {\bibfnamefont {S.}~\bibnamefont
  {de~Gironcoli}}, \bibinfo {author} {\bibfnamefont {P.}~\bibnamefont
  {Delugas}}, \bibinfo {author} {\bibfnamefont {R.~A.}\ \bibnamefont
  {DiStasio}}, \bibinfo {author} {\bibfnamefont {A.}~\bibnamefont {Ferretti}},
  \bibinfo {author} {\bibfnamefont {A.}~\bibnamefont {Floris}}, \bibinfo
  {author} {\bibfnamefont {G.}~\bibnamefont {Fratesi}}, \bibinfo {author}
  {\bibfnamefont {G.}~\bibnamefont {Fugallo}}, \bibinfo {author} {\bibfnamefont
  {R.}~\bibnamefont {Gebauer}}, \bibinfo {author} {\bibfnamefont
  {U.}~\bibnamefont {Gerstmann}}, \bibinfo {author} {\bibfnamefont
  {F.}~\bibnamefont {Giustino}}, \bibinfo {author} {\bibfnamefont
  {T.}~\bibnamefont {Gorni}}, \bibinfo {author} {\bibfnamefont
  {J.}~\bibnamefont {Jia}}, \bibinfo {author} {\bibfnamefont {M.}~\bibnamefont
  {Kawamura}}, \bibinfo {author} {\bibfnamefont {H.-Y.}\ \bibnamefont {Ko}},
  \bibinfo {author} {\bibfnamefont {A.}~\bibnamefont {Kokalj}}, \bibinfo
  {author} {\bibfnamefont {E.}~\bibnamefont {Küçükbenli}}, \bibinfo {author}
  {\bibfnamefont {M.}~\bibnamefont {Lazzeri}}, \bibinfo {author} {\bibfnamefont
  {M.}~\bibnamefont {Marsili}}, \bibinfo {author} {\bibfnamefont
  {N.}~\bibnamefont {Marzari}}, \bibinfo {author} {\bibfnamefont
  {F.}~\bibnamefont {Mauri}}, \bibinfo {author} {\bibfnamefont {N.~L.}\
  \bibnamefont {Nguyen}}, \bibinfo {author} {\bibfnamefont {H.-V.}\
  \bibnamefont {Nguyen}}, \bibinfo {author} {\bibfnamefont {A.~O.}\
  \bibnamefont {de-la Roza}}, \bibinfo {author} {\bibfnamefont
  {L.}~\bibnamefont {Paulatto}}, \bibinfo {author} {\bibfnamefont
  {S.}~\bibnamefont {Poncé}}, \bibinfo {author} {\bibfnamefont
  {D.}~\bibnamefont {Rocca}}, \bibinfo {author} {\bibfnamefont
  {R.}~\bibnamefont {Sabatini}}, \bibinfo {author} {\bibfnamefont
  {B.}~\bibnamefont {Santra}}, \bibinfo {author} {\bibfnamefont
  {M.}~\bibnamefont {Schlipf}}, \bibinfo {author} {\bibfnamefont {A.~P.}\
  \bibnamefont {Seitsonen}}, \bibinfo {author} {\bibfnamefont {A.}~\bibnamefont
  {Smogunov}}, \bibinfo {author} {\bibfnamefont {I.}~\bibnamefont {Timrov}},
  \bibinfo {author} {\bibfnamefont {T.}~\bibnamefont {Thonhauser}}, \bibinfo
  {author} {\bibfnamefont {P.}~\bibnamefont {Umari}}, \bibinfo {author}
  {\bibfnamefont {N.}~\bibnamefont {Vast}}, \bibinfo {author} {\bibfnamefont
  {X.}~\bibnamefont {Wu}},\ and\ \bibinfo {author} {\bibfnamefont
  {S.}~\bibnamefont {Baroni}},\ }\bibfield  {title} {\bibinfo {title} {Advanced
  capabilities for materials modelling with quantum espresso},\ }\href
  {https://doi.org/10.1088/1361-648X/aa8f79} {\bibfield  {journal} {\bibinfo
  {journal} {Journal of Physics: Condensed Matter}\ }\textbf {\bibinfo {volume}
  {29}},\ \bibinfo {pages} {465901} (\bibinfo {year} {2017})}\BibitemShut
  {NoStop}%
\bibitem [{\citenamefont {Perdew}\ \emph {et~al.}(1996)\citenamefont {Perdew},
  \citenamefont {Burke},\ and\ \citenamefont {Ernzerhof}}]{PBE}%
  \BibitemOpen
  \bibfield  {author} {\bibinfo {author} {\bibfnamefont {J.~P.}\ \bibnamefont
  {Perdew}}, \bibinfo {author} {\bibfnamefont {K.}~\bibnamefont {Burke}},\ and\
  \bibinfo {author} {\bibfnamefont {M.}~\bibnamefont {Ernzerhof}},\ }\bibfield
  {title} {\bibinfo {title} {Generalized gradient approximation made simple},\
  }\href {https://doi.org/10.1103/PhysRevLett.77.3865} {\bibfield  {journal}
  {\bibinfo  {journal} {Phys. Rev. Lett.}\ }\textbf {\bibinfo {volume} {77}},\
  \bibinfo {pages} {3865} (\bibinfo {year} {1996})}\BibitemShut {NoStop}%
\bibitem [{\citenamefont {Slater}\ and\ \citenamefont
  {Koster}(1954)}]{SlaterKoster}%
  \BibitemOpen
  \bibfield  {author} {\bibinfo {author} {\bibfnamefont {J.~C.}\ \bibnamefont
  {Slater}}\ and\ \bibinfo {author} {\bibfnamefont {G.~F.}\ \bibnamefont
  {Koster}},\ }\bibfield  {title} {\bibinfo {title} {Simplified lcao method for
  the periodic potential problem},\ }\href
  {https://doi.org/10.1103/PhysRev.94.1498} {\bibfield  {journal} {\bibinfo
  {journal} {Phys. Rev.}\ }\textbf {\bibinfo {volume} {94}},\ \bibinfo {pages}
  {1498} (\bibinfo {year} {1954})}\BibitemShut {NoStop}%
\bibitem [{sup()}]{supp}%
  \BibitemOpen
  \href@noop {} {}\bibinfo {howpublished}
  {\url{URL_will_be_inserted_by_publisher}}\BibitemShut {NoStop}%
\bibitem [{\citenamefont {Zhang}\ \emph {et~al.}(2022)\citenamefont {Zhang},
  \citenamefont {Liu}, \citenamefont {Deng}, \citenamefont {Shi}, \citenamefont
  {Tang}, \citenamefont {Chen},\ and\ \citenamefont {Yuan}}]{D2RA02988J}%
  \BibitemOpen
  \bibfield  {author} {\bibinfo {author} {\bibfnamefont {Y.}~\bibnamefont
  {Zhang}}, \bibinfo {author} {\bibfnamefont {J.}~\bibnamefont {Liu}}, \bibinfo
  {author} {\bibfnamefont {R.}~\bibnamefont {Deng}}, \bibinfo {author}
  {\bibfnamefont {X.}~\bibnamefont {Shi}}, \bibinfo {author} {\bibfnamefont
  {H.}~\bibnamefont {Tang}}, \bibinfo {author} {\bibfnamefont {H.}~\bibnamefont
  {Chen}},\ and\ \bibinfo {author} {\bibfnamefont {H.}~\bibnamefont {Yuan}},\
  }\bibfield  {title} {\bibinfo {title} {Electronic structure{,}
  magnetoresistance and spin filtering in graphene|2 monolayer-{CrI3}|graphene
  van der waals magnetic tunnel junctions},\ }\href
  {https://doi.org/10.1039/D2RA02988J} {\bibfield  {journal} {\bibinfo
  {journal} {RSC Adv.}\ }\textbf {\bibinfo {volume} {12}},\ \bibinfo {pages}
  {28533} (\bibinfo {year} {2022})}\BibitemShut {NoStop}%
\bibitem [{\citenamefont {Jiang}\ \emph {et~al.}(2018)\citenamefont {Jiang},
  \citenamefont {Li}, \citenamefont {Wang}, \citenamefont {Mak},\ and\
  \citenamefont {Shan}}]{dielectric3}%
  \BibitemOpen
  \bibfield  {author} {\bibinfo {author} {\bibfnamefont {S.}~\bibnamefont
  {Jiang}}, \bibinfo {author} {\bibfnamefont {L.}~\bibnamefont {Li}}, \bibinfo
  {author} {\bibfnamefont {Z.}~\bibnamefont {Wang}}, \bibinfo {author}
  {\bibfnamefont {K.~F.}\ \bibnamefont {Mak}},\ and\ \bibinfo {author}
  {\bibfnamefont {J.}~\bibnamefont {Shan}},\ }\bibfield  {title} {\bibinfo
  {title} {Controlling magnetism in 2d cri3 by electrostatic doping},\ }\href
  {https://doi.org/10.1038/s41565-018-0135-x} {\bibfield  {journal} {\bibinfo
  {journal} {Nature Nanotechnology}\ }\textbf {\bibinfo {volume} {13}},\
  \bibinfo {pages} {549} (\bibinfo {year} {2018})}\BibitemShut {NoStop}%
\bibitem [{\citenamefont {Jiang}\ \emph {et~al.}(2019)\citenamefont {Jiang},
  \citenamefont {Li}, \citenamefont {Wang}, \citenamefont {Shan},\ and\
  \citenamefont {Mak}}]{Jiang2019}%
  \BibitemOpen
  \bibfield  {author} {\bibinfo {author} {\bibfnamefont {S.}~\bibnamefont
  {Jiang}}, \bibinfo {author} {\bibfnamefont {L.}~\bibnamefont {Li}}, \bibinfo
  {author} {\bibfnamefont {Z.}~\bibnamefont {Wang}}, \bibinfo {author}
  {\bibfnamefont {J.}~\bibnamefont {Shan}},\ and\ \bibinfo {author}
  {\bibfnamefont {K.~F.}\ \bibnamefont {Mak}},\ }\bibfield  {title} {\bibinfo
  {title} {Spin tunnel field-effect transistors based on two-dimensional van
  der waals heterostructures},\ }\href
  {https://doi.org/10.1038/s41928-019-0232-3} {\bibfield  {journal} {\bibinfo
  {journal} {Nature Electronics}\ }\textbf {\bibinfo {volume} {2}},\ \bibinfo
  {pages} {159} (\bibinfo {year} {2019})}\BibitemShut {NoStop}%
\bibitem [{\citenamefont {Slizovskiy}\ \emph {et~al.}(2021)\citenamefont
  {Slizovskiy}, \citenamefont {Garcia-Ruiz}, \citenamefont {Berdyugin},
  \citenamefont {Xin}, \citenamefont {Taniguchi}, \citenamefont {Watanabe},
  \citenamefont {Geim}, \citenamefont {Drummond},\ and\ \citenamefont
  {Fal’ko}}]{Sliz1}%
  \BibitemOpen
  \bibfield  {author} {\bibinfo {author} {\bibfnamefont {S.}~\bibnamefont
  {Slizovskiy}}, \bibinfo {author} {\bibfnamefont {A.}~\bibnamefont
  {Garcia-Ruiz}}, \bibinfo {author} {\bibfnamefont {A.~I.}\ \bibnamefont
  {Berdyugin}}, \bibinfo {author} {\bibfnamefont {N.}~\bibnamefont {Xin}},
  \bibinfo {author} {\bibfnamefont {T.}~\bibnamefont {Taniguchi}}, \bibinfo
  {author} {\bibfnamefont {K.}~\bibnamefont {Watanabe}}, \bibinfo {author}
  {\bibfnamefont {A.~K.}\ \bibnamefont {Geim}}, \bibinfo {author}
  {\bibfnamefont {N.~D.}\ \bibnamefont {Drummond}},\ and\ \bibinfo {author}
  {\bibfnamefont {V.~I.}\ \bibnamefont {Fal’ko}},\ }\bibfield  {title}
  {\bibinfo {title} {Out-of-plane dielectric susceptibility of graphene in
  twistronic and bernal bilayers},\ }\href
  {https://doi.org/10.1021/acs.nanolett.1c02211} {\bibfield  {journal}
  {\bibinfo  {journal} {Nano Letters}\ }\textbf {\bibinfo {volume} {21}},\
  \bibinfo {pages} {6678} (\bibinfo {year} {2021})},\ \bibinfo {note} {pMID:
  34296602},\ \Eprint
  {https://arxiv.org/abs/https://doi.org/10.1021/acs.nanolett.1c02211}
  {https://doi.org/10.1021/acs.nanolett.1c02211} \BibitemShut {NoStop}%
\bibitem [{\citenamefont {Farooq}\ and\ \citenamefont
  {Hong}(2019)}]{Farooq2019}%
  \BibitemOpen
  \bibfield  {author} {\bibinfo {author} {\bibfnamefont {M.~U.}\ \bibnamefont
  {Farooq}}\ and\ \bibinfo {author} {\bibfnamefont {J.}~\bibnamefont {Hong}},\
  }\bibfield  {title} {\bibinfo {title} {Switchable valley splitting by
  external electric field effect in graphene/cri3 heterostructures},\ }\href
  {https://doi.org/10.1038/s41699-019-0086-6} {\bibfield  {journal} {\bibinfo
  {journal} {npj 2D Materials and Applications}\ }\textbf {\bibinfo {volume}
  {3}},\ \bibinfo {pages} {3} (\bibinfo {year} {2019})}\BibitemShut {NoStop}%
\bibitem [{\citenamefont {Laturia}\ \emph {et~al.}(2018)\citenamefont
  {Laturia}, \citenamefont {Van~de Put},\ and\ \citenamefont
  {Vandenberghe}}]{Laturia2018}%
  \BibitemOpen
  \bibfield  {author} {\bibinfo {author} {\bibfnamefont {A.}~\bibnamefont
  {Laturia}}, \bibinfo {author} {\bibfnamefont {M.~L.}\ \bibnamefont {Van~de
  Put}},\ and\ \bibinfo {author} {\bibfnamefont {W.~G.}\ \bibnamefont
  {Vandenberghe}},\ }\bibfield  {title} {\bibinfo {title} {Dielectric
  properties of hexagonal boron nitride and transition metal dichalcogenides:
  from monolayer to bulk},\ }\href {https://doi.org/10.1038/s41699-018-0050-x}
  {\bibfield  {journal} {\bibinfo  {journal} {npj 2D Materials and
  Applications}\ }\textbf {\bibinfo {volume} {2}},\ \bibinfo {pages} {6}
  (\bibinfo {year} {2018})}\BibitemShut {NoStop}%
\bibitem [{\citenamefont {Pierret}\ \emph {et~al.}(2022)\citenamefont
  {Pierret}, \citenamefont {Mele}, \citenamefont {Graef}, \citenamefont
  {Palomo}, \citenamefont {Taniguchi}, \citenamefont {Watanabe}, \citenamefont
  {Li}, \citenamefont {Toury}, \citenamefont {Journet}, \citenamefont {Steyer},
  \citenamefont {Garnier}, \citenamefont {Loiseau}, \citenamefont {Berroir},
  \citenamefont {Bocquillon}, \citenamefont {Fève}, \citenamefont {Voisin},
  \citenamefont {Baudin}, \citenamefont {Rosticher},\ and\ \citenamefont
  {Plaçais}}]{Pierret_2022}%
  \BibitemOpen
  \bibfield  {author} {\bibinfo {author} {\bibfnamefont {A.}~\bibnamefont
  {Pierret}}, \bibinfo {author} {\bibfnamefont {D.}~\bibnamefont {Mele}},
  \bibinfo {author} {\bibfnamefont {H.}~\bibnamefont {Graef}}, \bibinfo
  {author} {\bibfnamefont {J.}~\bibnamefont {Palomo}}, \bibinfo {author}
  {\bibfnamefont {T.}~\bibnamefont {Taniguchi}}, \bibinfo {author}
  {\bibfnamefont {K.}~\bibnamefont {Watanabe}}, \bibinfo {author}
  {\bibfnamefont {Y.}~\bibnamefont {Li}}, \bibinfo {author} {\bibfnamefont
  {B.}~\bibnamefont {Toury}}, \bibinfo {author} {\bibfnamefont
  {C.}~\bibnamefont {Journet}}, \bibinfo {author} {\bibfnamefont
  {P.}~\bibnamefont {Steyer}}, \bibinfo {author} {\bibfnamefont
  {V.}~\bibnamefont {Garnier}}, \bibinfo {author} {\bibfnamefont
  {A.}~\bibnamefont {Loiseau}}, \bibinfo {author} {\bibfnamefont {J.-M.}\
  \bibnamefont {Berroir}}, \bibinfo {author} {\bibfnamefont {E.}~\bibnamefont
  {Bocquillon}}, \bibinfo {author} {\bibfnamefont {G.}~\bibnamefont {Fève}},
  \bibinfo {author} {\bibfnamefont {C.}~\bibnamefont {Voisin}}, \bibinfo
  {author} {\bibfnamefont {E.}~\bibnamefont {Baudin}}, \bibinfo {author}
  {\bibfnamefont {M.}~\bibnamefont {Rosticher}},\ and\ \bibinfo {author}
  {\bibfnamefont {B.}~\bibnamefont {Plaçais}},\ }\bibfield  {title} {\bibinfo
  {title} {Dielectric permittivity, conductivity and breakdown field of
  hexagonal boron nitride},\ }\href {https://doi.org/10.1088/2053-1591/ac4fe1}
  {\bibfield  {journal} {\bibinfo  {journal} {Materials Research Express}\
  }\textbf {\bibinfo {volume} {9}},\ \bibinfo {pages} {065901} (\bibinfo {year}
  {2022})}\BibitemShut {NoStop}%
\bibitem [{\citenamefont {Ohba}\ \emph {et~al.}(2001)\citenamefont {Ohba},
  \citenamefont {Miwa}, \citenamefont {Nagasako},\ and\ \citenamefont
  {Fukumoto}}]{PhysRevB.63.115207}%
  \BibitemOpen
  \bibfield  {author} {\bibinfo {author} {\bibfnamefont {N.}~\bibnamefont
  {Ohba}}, \bibinfo {author} {\bibfnamefont {K.}~\bibnamefont {Miwa}}, \bibinfo
  {author} {\bibfnamefont {N.}~\bibnamefont {Nagasako}},\ and\ \bibinfo
  {author} {\bibfnamefont {A.}~\bibnamefont {Fukumoto}},\ }\bibfield  {title}
  {\bibinfo {title} {First-principles study on structural, dielectric, and
  dynamical properties for three bn polytypes},\ }\href
  {https://doi.org/10.1103/PhysRevB.63.115207} {\bibfield  {journal} {\bibinfo
  {journal} {Phys. Rev. B}\ }\textbf {\bibinfo {volume} {63}},\ \bibinfo
  {pages} {115207} (\bibinfo {year} {2001})}\BibitemShut {NoStop}%
\bibitem [{\citenamefont {Zhang}\ \emph {et~al.}(2021)\citenamefont {Zhang},
  \citenamefont {Guo}, \citenamefont {Li}, \citenamefont {Wang}, \citenamefont
  {Zhou}, \citenamefont {Zhao}, \citenamefont {Guo},\ and\ \citenamefont
  {Zhong}}]{doi:10.1021/acs.jpclett.1c00112}%
  \BibitemOpen
  \bibfield  {author} {\bibinfo {author} {\bibfnamefont {J.}~\bibnamefont
  {Zhang}}, \bibinfo {author} {\bibfnamefont {Y.}~\bibnamefont {Guo}}, \bibinfo
  {author} {\bibfnamefont {P.}~\bibnamefont {Li}}, \bibinfo {author}
  {\bibfnamefont {J.}~\bibnamefont {Wang}}, \bibinfo {author} {\bibfnamefont
  {S.}~\bibnamefont {Zhou}}, \bibinfo {author} {\bibfnamefont {J.}~\bibnamefont
  {Zhao}}, \bibinfo {author} {\bibfnamefont {D.}~\bibnamefont {Guo}},\ and\
  \bibinfo {author} {\bibfnamefont {D.}~\bibnamefont {Zhong}},\ }\bibfield
  {title} {\bibinfo {title} {Imaging vacancy defects in single-layer chromium
  triiodide},\ }\href {https://doi.org/10.1021/acs.jpclett.1c00112} {\bibfield
  {journal} {\bibinfo  {journal} {The Journal of Physical Chemistry Letters}\
  }\textbf {\bibinfo {volume} {12}},\ \bibinfo {pages} {2199} (\bibinfo {year}
  {2021})},\ \bibinfo {note} {pMID: 33630596},\ \Eprint
  {https://arxiv.org/abs/https://doi.org/10.1021/acs.jpclett.1c00112}
  {https://doi.org/10.1021/acs.jpclett.1c00112} \BibitemShut {NoStop}%
\bibitem [{\citenamefont {Magorrian}\ \emph {et~al.}(2016)\citenamefont
  {Magorrian}, \citenamefont {Z\'olyomi},\ and\ \citenamefont
  {Fal'ko}}]{PhysRevB.94.245431}%
  \BibitemOpen
  \bibfield  {author} {\bibinfo {author} {\bibfnamefont {S.~J.}\ \bibnamefont
  {Magorrian}}, \bibinfo {author} {\bibfnamefont {V.}~\bibnamefont
  {Z\'olyomi}},\ and\ \bibinfo {author} {\bibfnamefont {V.~I.}\ \bibnamefont
  {Fal'ko}},\ }\bibfield  {title} {\bibinfo {title} {Electronic and optical
  properties of two-dimensional inse from a dft-parametrized tight-binding
  model},\ }\href {https://doi.org/10.1103/PhysRevB.94.245431} {\bibfield
  {journal} {\bibinfo  {journal} {Phys. Rev. B}\ }\textbf {\bibinfo {volume}
  {94}},\ \bibinfo {pages} {245431} (\bibinfo {year} {2016})}\BibitemShut
  {NoStop}%
\bibitem [{\citenamefont {Bandurin}\ \emph {et~al.}(2017)\citenamefont
  {Bandurin}, \citenamefont {Tyurnina}, \citenamefont {Yu}, \citenamefont
  {Mishchenko}, \citenamefont {Z{\'o}lyomi}, \citenamefont {Morozov},
  \citenamefont {Kumar}, \citenamefont {Gorbachev}, \citenamefont {Kudrynskyi},
  \citenamefont {Pezzini}, \citenamefont {Kovalyuk}, \citenamefont {Zeitler},
  \citenamefont {Novoselov}, \citenamefont {Patan{\`e}}, \citenamefont {Eaves},
  \citenamefont {Grigorieva}, \citenamefont {Fal'ko}, \citenamefont {Geim},\
  and\ \citenamefont {Cao}}]{Bandurin2017}%
  \BibitemOpen
  \bibfield  {author} {\bibinfo {author} {\bibfnamefont {D.~A.}\ \bibnamefont
  {Bandurin}}, \bibinfo {author} {\bibfnamefont {A.~V.}\ \bibnamefont
  {Tyurnina}}, \bibinfo {author} {\bibfnamefont {G.~L.}\ \bibnamefont {Yu}},
  \bibinfo {author} {\bibfnamefont {A.}~\bibnamefont {Mishchenko}}, \bibinfo
  {author} {\bibfnamefont {V.}~\bibnamefont {Z{\'o}lyomi}}, \bibinfo {author}
  {\bibfnamefont {S.~V.}\ \bibnamefont {Morozov}}, \bibinfo {author}
  {\bibfnamefont {R.~K.}\ \bibnamefont {Kumar}}, \bibinfo {author}
  {\bibfnamefont {R.~V.}\ \bibnamefont {Gorbachev}}, \bibinfo {author}
  {\bibfnamefont {Z.~R.}\ \bibnamefont {Kudrynskyi}}, \bibinfo {author}
  {\bibfnamefont {S.}~\bibnamefont {Pezzini}}, \bibinfo {author} {\bibfnamefont
  {Z.~D.}\ \bibnamefont {Kovalyuk}}, \bibinfo {author} {\bibfnamefont
  {U.}~\bibnamefont {Zeitler}}, \bibinfo {author} {\bibfnamefont {K.~S.}\
  \bibnamefont {Novoselov}}, \bibinfo {author} {\bibfnamefont {A.}~\bibnamefont
  {Patan{\`e}}}, \bibinfo {author} {\bibfnamefont {L.}~\bibnamefont {Eaves}},
  \bibinfo {author} {\bibfnamefont {I.~V.}\ \bibnamefont {Grigorieva}},
  \bibinfo {author} {\bibfnamefont {V.~I.}\ \bibnamefont {Fal'ko}}, \bibinfo
  {author} {\bibfnamefont {A.~K.}\ \bibnamefont {Geim}},\ and\ \bibinfo
  {author} {\bibfnamefont {Y.}~\bibnamefont {Cao}},\ }\bibfield  {title}
  {\bibinfo {title} {High electron mobility, quantum hall effect and anomalous
  optical response in atomically thin inse},\ }\href
  {https://doi.org/10.1038/nnano.2016.242} {\bibfield  {journal} {\bibinfo
  {journal} {Nature Nanotechnology}\ }\textbf {\bibinfo {volume} {12}},\
  \bibinfo {pages} {223} (\bibinfo {year} {2017})}\BibitemShut {NoStop}%
\end{thebibliography}%
	
\end{document}